\documentclass[usegraphicx,usenatbib]{mn2e}
\usepackage{amsmath}
\usepackage{amssymb}
\usepackage{color}

\pdfoutput=1
\usepackage{threeparttable}

\newcommand{\lya}{\ifmmode\mathrm{Ly}\alpha\else{}Ly$\alpha$\fi}
\newcommand{\lyb}{\ifmmode\mathrm{Ly}\beta\else{}Ly$\beta$\fi}
\newcommand{\igm}{\ifmmode\mathrm{IGM}\else{}IGM\fi}
\newcommand{\lae}{\ifmmode\mathrm{LAE}\else{}LAE\fi}
\newcommand{\bao}{\ifmmode\mathrm{BAO}\else{}BAO\fi}
\newcommand{\h}{\ifmmode\mathrm{H}\else{}H\fi}
\newcommand{\hi}{\ifmmode\mathrm{H\,{\scriptscriptstyle I}}\else{}H\,{\scriptsize I}\fi}
\newcommand{\hii}{\ifmmode\mathrm{H\,{\scriptscriptstyle II}}\else{}H\,{\scriptsize II}\fi}
\newcommand{\he}{\ifmmode\mathrm{He}\else{}He\fi}
\newcommand{\hei}{\ifmmode\mathrm{He\,{\scriptscriptstyle I}}\else{}He\,{\scriptsize I}\fi}
\newcommand{\heii}{\ifmmode\mathrm{He\,{\scriptscriptstyle II}}\else{}He\,{\scriptsize II}\fi}
\newcommand{\heiii}{\ifmmode\mathrm{He\,{\scriptscriptstyle III}}\else{}He\,{\scriptsize III}\fi}
\newcommand{\ciii}{\ifmmode\mathrm{C\,{\scriptscriptstyle III]}}\else{}C\,{\scriptsize III]}\fi}
\newcommand{\oiii}{\ifmmode\mathrm{O\,{\scriptscriptstyle III}}\else{}O\,{\scriptsize III}\fi}
\newcommand{\aliii}{\ifmmode\mathrm{Al\,{\scriptscriptstyle III}}\else{}Al\,{\scriptsize III}\fi}
\newcommand{\mgii}{\ifmmode\mathrm{Mg\,{\scriptscriptstyle II}}\else{}Mg\,{\scriptsize II}\fi}
\newcommand{\fe}{\ifmmode\mathrm{Fe}\else{}Fe\fi}
\newcommand{\cmb}{\ifmmode\mathrm{CMB}\else{}CMB\fi}
\newcommand{\qso}{\ifmmode\mathrm{QSO}\else{}QSO\fi}
\newcommand{\nv}{\ifmmode\mathrm{N\,{\scriptscriptstyle V}}\else{}N\,{\scriptsize V}\fi}
\newcommand{\niv}{\ifmmode\mathrm{N\,{\scriptscriptstyle IV]}}\else{}N\,{\scriptsize IV]}\fi}
\newcommand{\civ}{\ifmmode\mathrm{C\,{\scriptscriptstyle IV}}\else{}C\,{\scriptsize IV}\fi}
\newcommand{\siv}{\ifmmode\mathrm{Si\,{\scriptscriptstyle IV}}\else{}Si\,{\scriptsize IV}\fi}
\newcommand{\oiv}{\ifmmode\mathrm{O\,{\scriptscriptstyle IV]}}\else{}O\,{\scriptsize IV]}\fi}
\newcommand{\siii}{\ifmmode\mathrm{Si\,{\scriptscriptstyle II}}\else{}Si\,{\scriptsize II}\fi}
\newcommand{\ovi}{\ifmmode\mathrm{O\,{\scriptscriptstyle VI}}\else{}O\,{\scriptsize VI}\fi}
\newcommand{\sioiv}{\ifmmode\mathrm{Si\,{\scriptscriptstyle IV}\,\plus O\,{\scriptscriptstyle IV]}}\else{}Si\,{\scriptsize IV}\,+O\,{\scriptsize IV]}\fi}

\title[\lya{} emission line reconstruction for high-$z$ QSOs]{\lya{} emission line reconstruction for high-$z$ QSOs}

\author[B. Greig et al.] {Bradley~Greig$^{1}$\thanks{E-mail:~bradley.greig@sns.it~(BG),}, Andrei~Mesinger$^{1}$, Ian D. McGreer$^{2}$, Simona Gallerani$^{1}$, \newauthor Zolt{\'a}n~Haiman$^{3}$\\
$^1$Scuola Normale Superiore, Piazza dei Cavalieri 7, I-56126 Pisa, Italy \\
$^2$Steward Observatory, The University of Arizona, 933 North Cherry Avenue, Tucson, AZ 85721-0065, USA \\
$^3$Department of Astronomy, Columbia University, 550 West 120th Street, New York, NY 10027, USA
}

\begin{document}
\maketitle \begin{abstract}
\noindent
We introduce an intrinsic \lya{} emission line profile reconstruction method for high-$z$ quasars (QSOs). This approach utilises a covariance matrix of emission line properties obtained from a large, moderate-$z$ ($2 \leq z \leq 2.5$), high signal to noise (S/N$>15$) sample of BOSS QSOs. For each QSO, we complete a Monte Carlo Markov Chain fitting of the continuum and emission line properties and perform a visual quality assessment to construct a large database of robustly fit spectra. With this dataset, we construct a covariance matrix to describe the correlations between the high ionisation emission lines \lya{}, \civ{}, \sioiv{} and \ciii{}, and find it to be well approximated by an $N$-dimensional Gaussian distribution. This covariance matrix characterises the correlations between the line width, peak height and velocity offset from systemic while also allowing for the existence of broad and narrow line components for \lya{} and \civ{}. We illustrate how this covariance matrix allows us to statistically characterise the intrinsic \lya{} line solely from the observed spectrum redward of 1275\AA. This procedure can be used to reconstruct the intrinsic \lya{} line emission profile in cases where \lya{} may otherwise be obscured. Applying this reconstruction method to our sample of QSOs, we recovered the \lya{} line flux to within 15 per cent of the measured flux at 1205\AA~(1220\AA)~$\sim85$ (90) per cent of the time.
\end{abstract} 
\begin{keywords}
quasars: emission lines -- quasars: absorption lines -- quasars: general -- cosmology: observations -- cosmology: theory
\end{keywords}

\section{Introduction}

The ability to measure the intrinsic spectrum of quasars (QSOs) plays an important role in astrophysics. Resolving individual emission line profiles can enable the detailed study of both the central supermassive black hole (SMBH) and the internal structure within the SMBH powered accretion disk (through reverberation mapping e.g.\ \citealt{Blandford:1982,Peterson:1993} or the virial method e.g.\ \citealt{Kaspi:2000,Peterson:2004,Vestergaard:2006}). On cosmological scales, the relative strength (or lack thereof) of the emission profiles and spectral slopes of the QSO continuum can be used to yield insights into the thermal and ionisation state of the intergalactic medium (IGM) through studies of the QSO proximity zone \citep[e.g.][]{Mesinger:2004p3625,Wyithe:2004,Fan:2006p4005,Bolton:2007p3623,Mesinger:2007p855,Carilli:2010p1,Calverley:2011,Wyithe:2011,Schroeder:2013p919}. 

The key ingredient for using QSOs to explore the IGM is observing the intrinsic ultraviolet (UV) emission. Thermal emission from the accretion disk peaks in UV, which then interacts with surrounding neutral hydrogen and recombines two thirds of the time into \lya{} photons (rest frame $\lambda_{\lya{}}=1215.67$\AA) to produce a prominent \lya{} emission profile. However, owing to the large scattering cross-section of \lya{} photons, neutral hydrogen column densities of $N_{\hi{}} > 10^{18} {\rm cm}^{-2}$ are sufficient for \lya{} to enter into the strong absorption regime. In the case of the IGM, even minute traces ($\sim$~few per cent) of intervening neutral hydrogen along the line-of-sight are capable of scattering \lya{} photons away from the observer.

At $z\lesssim3$, the IGM is on average very highly ionised except in dense self shielded clumps \citep[e.g.][]{Fan:2006p4005,McGreer:2015p3668,Collaboration:2015p4320,Collaboration:2016,Collaboration:2016p5913}. Typically, at these redshifts the resonant scattering and absorption of \lya{} photons occurs only due to diffuse amounts of neutral hydrogen which appear blueward of \lya{} ($\lambda<\lambda_{\lya{}}$) as a series of discrete narrow absorption features, known as the \lya{} forest \citep{Rauch:1998}. More problematic for measuring the intrinsic \lya{} emission line are larger neutral column density absorbers such as Lyman-limit systems and damped \lya{} absorbers (DLAs). In the case of DLAs, with column densities of $N_{\hi{}} > 10^{20} {\rm cm}^{-2}$ not only do these systems lead to completely saturated absorption, but they are also sufficiently dense to allow \lya{} absorption within the Lorentzian wings. If these DLAs are located sufficiently close to the source QSO, absorption in the wings can significantly affect the intrinsic \lya{} emission profile. Furthermore, associated strong absorbers intrinsic to the host QSO environment itself can additionally impact the \lya{} line emission \citep{Shen:2012}.

At $z>6$, the IGM becomes increasingly neutral, and the \lya{} forest gives way to completely dark (absorbed) patches \citep[e.g.][]{Barkana:2002,Gallerani:2008,Mesinger:2010p6068,McGreer:2015p3668}. Once the IGM itself obtains a sufficiently large column density, it too begins to absorb \lya{} in the Lorentzian wings of the scattering cross-section, referred to as IGM damping wing absorption \citep{MiraldaEscude:1998p1041}. In principle, through the detection of the IGM damping wing imprint from the \lya{} emission spectra of high-$z$ QSOs one is able to provide a direct measurement of the neutral fraction of the IGM during the reionisation epoch. Importantly, this approach requires an estimate of the intrinsic \lya{} emission line in order to determine the contribution from the smooth component absorption of the IGM damping wing. Using this technique, limits on the IGM neutral fraction at $z>6$ have been obtained \citep[e.g.][]{Mesinger:2007p855,Bolton:2011p1063,Schroeder:2013p919}. 

In the absence of a viable method to recover the intrinsic \lya{} emission profile, one can attempt to use an emission template constructed from a QSO composite spectrum. Numerous composite spectra exist, constructed by averaging over a vastly different number of QSOs \citep[e.g.][]{Francis:1991p5112,Brotherton:2001p1,VandenBerk:2001p3887,Telfer:2002p5713,Shull:2012p5716,Stevans:2014p5726,Harris:2016p5028}. However, by construction, these composite spectra only describe the average properties of QSOs, not the intrinsic variations of individual QSOs. Not accounting for these intrinsic properties can bias estimates of the IGM damping wing \citep{Mortlock:2011p1049,Bosman:2015p5005}.

An alternative method is to reconstruct a template using a principle component analysis (PCA) \citep[e.g.][]{Boroson:1992p4641,Francis:1992p5021,Suzuki:2005p5157,Suzuki:2006p4770,Lee:2011p1738,Paris:2011p4774}. An improvement over the use of a composite spectrum, this approach aims to use the minimal subset of eigenvectors to characterise the QSO emission profile. For example, \citet{Francis:1992p5021} find 3 eigenvectors are sufficient to describe 75 per cent of the observed profile variation, and 95 per cent if 10 eigenvectors are used. Beyond 10, the eigenvectors obtain little information and become rather noisy \citep{Suzuki:2005p5157}. However, these eigenvectors are extracted from a fit to the original QSO spectrum. Therefore, given that the \lya{} line cannot be directly recovered at high redshifts, in order to obtain an estimate of the intrinsic \lya{} profile some form of reconstruction/extrapolation of the relevant eigenvectors that describe \lya{} would be required.

Conveniently, studies of these QSO composites and PCA approaches have revealed a wealth of information regarding correlations amongst various emission lines and other observable properties of the source QSO. For example, the first eigenvector of \citet{Boroson:1992p4641} showed that in the H$\beta$ region a strong anti-correlation existed between [O\,{\scriptsize{III}}] and Fe\,{\scriptsize{II}}. Furthermore, \citet{Hewett:2010p6194} performed an in-depth exploration to establish relationships between emission line profiles and the source systemic redshift to reduce the scatter and biases in redshift determination. More directly relevant for this work, both \citet{Shang:2007p4862} and \citet{Kramer:2009p920} observe a strong correlation between the \lya{} and \civ{} peak blue shifts. 

Motivated by the existence of correlations amongst the various emission lines properties, and the lack of either a robust physical model of quasar emission regions \citep[e.g.][]{Baldwin:1995} or a method to recover the intrinsic \lya{} profile within high-$z$ or heavily \lya{} obscured QSOs, we propose a new method to reconstruct the intrinsic \lya{} emission line profile for any high-$z$ or \lya{} obscured QSO. In this work, we develop our reconstruction method using QSOs selected from the Baryon Oscillation Spectroscopic Survey (BOSS; \citealt{Dawson:2013p5160}), a component of SDSS-III \citep{Eisenstein:2011p5159}. In summary, our \lya{} reconstruction method is as follows:
\begin{itemize}
\item Perform a Monte Carlo Markov Chain (MCMC) fit to a subset of emission lines for each of our selected QSOs.
\item Construct a covariance matrix which describes all correlations amongst the QSO emission lines.
\item Assume a $N$-dimensional Gaussian likelihood function to describe the covariance matrix.
\item MCMC fit a high-$z$ or \lya{} obscured QSO characterising all lines except \lya{}.
\item Use the recovered emission line information from the QSO to statistically characterise the intrinsic \lya{} line.
\end{itemize}

The remainder of this paper is organised as follows. In Section~\ref{sec:Data}, we discuss the observational data used within this work, and the selection criteria we apply to construct our sample of QSOs. In Section~\ref{sec:Fitting} we describe our MCMC fitting procedure, and how we model the QSO continuum, emission line features and other components to aid the estimation of the observed QSO flux. With all the data obtained from fitting our entire QSO sample, in Section~\ref{sec:Covariance} we construct our covariance matrix, and discuss the major correlations and recovered features. In Section~\ref{sec:Recon} we outline our \lya{} reconstruction method, and highlight the performance of this approach. Following this, in Section~\ref{sec:Discussions} we provide a discussion of the potential applications for both the MCMC fitting algorithm and the \lya{} reconstruction pipeline. Finally, in Section~\ref{sec:Conclusions} we finish with our closing remarks.

\section{Data} \label{sec:Data}

In this work, we select our QSOs from Data Release 12 (DR12) \citep{Alam:2015p5162} of the large-scale SDSS-III observational programme BOSS \citep{Dawson:2013p5160}. The full details of the SDSS telescope are available in \citet{Gunn:2006p1} and the details of the upgraded SDSS/BOSS spectrographs may be obtained in \citet{Smee:2013p1}. For reference, the wavelength coverage of the BOSS spectrograph is $3,\!600{\rm \AA} < \lambda < 10,\!400{\rm \AA}$ in the observed frame, with a resolution of $R\sim2000-2500$ corresponding to pixel resolution of $\sim120-150$~km/s. The QSO target selection for BOSS consisted of a variety of schemes, including colour and variability selection, counterparts to radio and X-ray sources, and previously known QSOs \citep{Bovy:2011p1,Kirkpatrick:2011p1,Ross:2012p1}. The candidate QSO spectra were then visually inspected following the procedure outlined in \citet{Paris:2016p1}, with updated lists of confirmed QSOs for DR12 available online\footnote{http://www.sdss.org/dr12/algorithms/boss-dr12-quasar-catalog/}. Furthermore, we use the publicly available flux calibration model of \citet{Margala:2015p1} to perform the spectrophotometric corrections of the DR12 QSO fluxes.

In total, DR12, contains 294,512 uniquely identified QSOs, 158,917 of which are observed within the redshift range $2.15 < z < 3.5$. The principal science goal of the BOSS observational programme was the detection of the baryon acoustic oscillation scale from the \lya{} forest. To perform this, only relatively low to moderate S/N QSOs are required \citep[e.g.][]{White:2003p578,McDonald:2007p6752}. However, within this work, we aim to statistically characterise the correlations between numerous QSO emission lines, which require much higher S/N. 

To construct the QSO sample used within this work, we preferentially selected QSOs with a median S/N across all filters ($ugriz$) of S/N~$>$~15 (`snMedian~$>15$'). This choice is arbitrary, and in principle we could go lower, however for decreasing S/N the weaker emission lines become more difficult to differentiate from the noise, reducing potential correlations. We selected QSOs containing broad-line emission using the `BROADLINE' flag and removed all sources visually confirmed to contain broad absorption lines (BALs) by using the `QSO' selection flag. Furthermore, only QSOs with `ZWARNING' set to zero are retained, where the redshifts were recovered with high confidence from the BOSS pipeline. Finally, we restrict our QSO redshift range to $2.08 < z < 2.5$\footnote{This choice of redshift range was a trade off between the wavelength coverage of the spectrograph and requiring that we were sufficiently blueward of rest-frame \lya{} to characterise the line profile (arbitrarily chosen to be 1180\AA, corresponding to $\sim3600$\AA~at $z=2.08$) while also including the \mgii{} emission line ($\lambda=2798.75$\AA, corresponding to $\lambda = 10,000$\AA~at $z=2.5$).}, selecting QSOs by their BOSS pipeline redshift\footnote{Note that the r.m.s of the quasar redshift error distribution for the BOSS pipeline redshift from DR9 was determined to be $550$~km/s with a mean offset of the quasar redshift (bias) of $\sim-150$~km/s \citep{Font-Ribera:2013}.} (e.g.\ \citealt{Bolton:2012SDSS}; see Appendix~\ref{sec:systemic_redshift} for a more in-depth discussion on our adopted choice of BOSS redshift). Following these selection cuts, we recover a total QSO sample of 3,862\footnote{We computed the absolute AB magnitude from the QSO continuum at 1450\AA\,($M_{1450}$) for this sample of QSOs and recovered a median of $M_{1450} = -26.1$ with an interquartile range of 0.7.}. 

Though the total QSO number of 3,862 appears small, it should be more than sufficient to elucidate any statistically significant correlations amongst the emission line parameters\footnote{The errors and biases when attempting to accurately measure the covariance matrix of a set of parameters ($N_{\rm par}$) from a total number of sources ($N_{\rm src}$) scales approximately as $N_{\rm par}$/$N_{\rm src}$ \citep[e.g][]{Dodelson:2013,Taylor:2014,Petri:2016}. Within this work, we ultimately construct an 18 parameter covariance matrix from a subsample of 1673 QSOs, corresponding to an expected error of $\sim1$~per cent.}. Furthermore, to properly assess the MCMC fitting to be discussed in Section~\ref{sec:Fitting} a detailed visual inspection will be required, therefore it is preferential to restrict the sample size. In selecting our redshift range of $2.08 < z < 2.5$, we are assuming there is no strong variation in the emission line profiles of QSOs across the age of the Universe. That is, the covariance matrix recovered from this QSO sample will always be representative of the underlying QSO population. In \citet{Becker:2013p1008}, these authors constructed 26 QSO composite spectra between rest-frame $1040 < \lambda < 1550$\AA~ across the redshift range of $2 < z < 5$. They found for \lya{} and the emission lines redward, no significant variation with redshift, lending confidence to our assumption.

Note, throughout this work, we do not deredden our QSOs to account for interstellar dust extinction within the Milky Way \citep[e.g.][]{Fitzpatrick:1999p5034}. The dust extinction curve varies strongly in the UV and ultimately will impact the \lya{} line more than the \civ{} line. However, this will not greatly impact our results as we are attempting to reconstruct the \lya{} emission line using a covariance matrix of correlations from other emission lines. Applying the extinction curve will only impact slightly on the peak of the emission lines, not the other characteristics of the line profile. When highlighting the performance of the reconstruction process in Section~\ref{sec:Recon}, we see that this slight overestimate should be well within the errors of the reconstructed profile.

\section{MCMC QSO Fitting} \label{sec:Fitting}

We now introduce our MCMC fitting procedure. In Section~\ref{sec:lines}, we outline and justify the selection of the emission lines considered. In Section~\ref{sec:template} we then detail the construction of the QSO template. In order to improve the ability to recover the intrinsic emission line profile and QSO continuum we outline the treatment of `absorption' features in Section~\ref{sec:absorb}. In Section~\ref{sec:MCMC} the iterative MCMC fitting procedure is discussed and an example is presented. Finally, in Section~\ref{sec:QA} we perform a visual quality assessment of our entire sample of fit QSOs to remove contaminants that could impact the observed emission line correlations.

\subsection{Emission line selection} \label{sec:lines}

Owing to the limited wavelength coverage of BOSS, the selection of available emission lines to be used is limited. Due to the complexity in observing high-$z$ ($z\gtrsim6$) QSOs, where \lya{} is redshifted into the near infrared (IR), the emission lines we use from the BOSS sample needs to be consistent with what is detectable with near-IR instruments such as Keck/MOSFIRE \citep{McLean:2010,McLean:2012} and VLT/X-Shooter \citep{Vernet:2011}. 

The strongest emission lines, especially \lya{}, \civ{} and \mgii{}, are known to contain a broad and narrow component \citep[e.g.][]{Wills:1993,Baldwin:1996,VandenBerk:2001p3887,Shang:2007p4862,Kramer:2009p920,Shen:2011p4583}. These different components are thought to arise from moving clouds of gas above and below the accretion disk known as the broad and narrow line regions. Note that, for our purposes, the physical origins of these components is irrelevant; a double gaussian merely provides us a flexible basis set in which to characterise the line profile. Throughout this work, we will approximate all emission lines to have a Gaussian profile. While emission line profiles are known to be Lorentzian \citep[e.g.][]{Peebles1993}, the fitting of a Lorentzian profile is complicated by the requisite identification of the broader wings relative to the uncertainty in the QSO continuum\footnote{In practice, thermal (or other) broadening will make the Lorentzian wings irrelevant in the case of AGN emission lines. For even modest thermal broadening of $\sim10$~km/s only the Gaussian core can be detected \citep[see e.g.\ figure~4 of][]{Dijkstra:2014p1}.}. Ultimately, this is a subtle difference as we are only interested in characterising the total line profile.

In order to decide which strong emission line profiles should be fit with a single or double component Gaussian\footnote{In principle, we could allow for $N>2$ components to our line fitting procedure. For example, several authors have fitted three Gaussian components to the \civ{} emission line profile \citep[e.g.][]{Dietrich:2002p5912,Shen:2008p5975,Shen:2011p4583}. However, in this work, we attempt to avoid adding too much complexity to the fitting method.}, we perform a simple test analysing the Bayes information criteria (BIC; \citealt{Schwarz:1978p1,Liddle:2004p5730}) of each of the individual line profiles in Appendix~\ref{sec:line_component}. Below we summarise our findings:
\begin{itemize}
\item \lya{}: The strongest observed UV emission line. In order to better characterise the fit to the line profile, a two component Gaussian is preferred. This is well known and consistent with other works.
\item \sioiv{}: This line complex consists of both the \siv{} ($\lambda$1396.76\AA) and \oiv{} ($\lambda$1402.06\AA) line centres. Unfortunately, these individual line profiles are intrinsically broad, preventing our ability to distinguish between them. Therefore these two lines appear as a single, blended line. In Appendix~\ref{sec:line_component} we find a single Gaussian component is sufficient to characterise the line profile.
\item \civ{}: This line profile is a doublet ($\lambda$$\lambda$1548.20, 1550.78\AA). Again, both components being intrinsically broad prevents individual detection, therefore the line profile is observed as a single, strong line. For the \civ{} line, we find a two component Gaussian to be preferable to characterise the line.
\item \ciii{}: For the \ciii{} line ($\lambda$1908.73\AA) we find a single component Gaussian to be sufficient to characterise the line. For almost all of the BOSS QSOs in our sample, the nearby, weak Si\,{\scriptsize{III]}} line ($\lambda$1892.03\AA), which would appear on top of the much broader \ciii{} line is not resolvable\footnote{
In principle, by ignoring the Si\,{\scriptsize{III]}} feature, the centroid of the measured \ciii{} line can be shifted bluer than its true value. Quantitatively, we determine the extent of this blueshift by fitting all QSOs within our sample that have a clear, discernible Si\,{\scriptsize{III]}} feature. We find that by excluding Si\,{\scriptsize{III]}} the average recovered \ciii{} emission line centre is blue shifted by $\sim310$~km/s (with the associated line width broadened by $\sim315$~km/s). This sample, however, only contains 57 (out of 1673) QSOs, corresponding to only $\sim3.4$ per cent of the full `good' sample affected by this blueshift. For the vast majority of QSOs in our sample the S/N around \ciii{} is insufficient to tease out the individual emission component of the Si\,{\scriptsize{III]}} feature. Therefore, owing to the small resultant blueshift in the line centre estimate (compared to the average dispersion of $\sim550$~km/s in the redshift estimate of the chosen BOSS pipeline redshift) and the corresponding low number of sources potentially affected by this systematic shift, we deem the exclusion of Si\,{\scriptsize{III]}} a valid simplification. As a final point, owing to the weak total flux expected in the Si\,{\scriptsize{III]}} emission line, including this in the MCMC fitting for the full QSO sample (where Si\,{\scriptsize{III]}} is not clearly present) would cause a strong degeneracy with the \ciii{} line unless extremely strong priors on the line shape are imposed.}. Furthermore, the \ciii{} line may also be contaminated by a continuum of low ionisation \fe{} lines, however, unlike \mgii{} (see next) the relative impact on the \ciii{} line is minor.
\item \mgii{}: While the \mgii{} line ($\lambda$2798.75\AA) is observable within the wavelength range of our BOSS QSO sample, we choose not to fit this emission profile. This is because of the contamination from the \fe{} pseudo-continuum. Though \fe{} emission templates exist \citep[e.g.][]{Vestergaard:2001p3921}, the fitting procedure is complicated. One must simultaneously fit several spectral regions in order to calibrate the true flux level of the \fe{} pseudo-continuum in order to remove it and obtain an estimate of the \mgii{} emission line. Since we aim to simultaneously fit the full QSO spectrum, this approach would add degeneracies to the model (e.g.\ between the true continuum, and the \fe{} pseudo-continuum). While this can be overcome, it requires first fitting the QSO continuum blueward of \mgii{}, before fitting the \fe{} pseudo-continuum, which reduces the flexibility of our model.
\end{itemize}

In addition to these lines above, we also consider several other lines all modelled by a single Gaussian. These include: (i) \nv{} ($\lambda$$\lambda$1238.8, 1242.8\AA), which is an important line for characterising the \lya{} line profile as it can be degenerate with the broad \lya{} line component (ii) \siii{} ($\lambda$1262.59\AA), (iii) the O\,{\scriptsize I}/Si\,{\scriptsize II} blended complex ($\lambda$1304.35, $\lambda$1306.82\AA), (iv) C\,{\scriptsize II} ($\lambda$1335.30\AA), (v) \heii{} ($\lambda$1640.42\AA) (vi) O\,{\scriptsize III} ($\lambda$1663.48\AA) and (vii) Al\,{\scriptsize III} ($\lambda$1857.40\AA). 

\subsection{Continuum and emission line template} \label{sec:template}

In this work, we fit the rest-frame wavelength range ($1180{\rm \AA} < \lambda < 2300{\rm \AA}$) with a single power-law for the QSO continuum,
\begin{eqnarray} \label{eq:continuum}
f_{\lambda} = f_{1450}\left(\frac{\lambda}{1450{\rm \AA}}\right)^{\alpha_{\lambda}} {\rm erg\,cm^{-2}\,s^{-1}\,\AA^{-1}},
\end{eqnarray}
where $\alpha_{\lambda}$ describes the spectral slope of the continuum and $f_{1450}$ is the normalisation of the QSO flux which we choose to measure at 1450\AA. While we normalise the QSO flux at 1450\AA, we allow this quantity to vary within our MCMC fitting algorithm (by adding a small, variable perturbation). While this quantity does not depart greatly from the original normalised value, it allows us to compensate for situations where the nearby region around 1450\AA\, might be impacted by line absorption or a noise feature from the spectrograph.

Secondary (broken) power-law continua have been fit to QSO spectra at $\lambda > 4500$\AA~for an independent red continuum slope \citep[e.g.][]{VandenBerk:2001p3887,Shang:2007p4862}, however, this is beyond our QSO sample wavelength coverage. Typically, a broken power-law continuum is also adopted blueward of \lya{}, visible in low-$z$ HST spectra \citep[e.g][]{Telfer:2002p5713,Shull:2012p5716}. In this work, we do not consider a different slope near \lya{} as we only fit down to 1180\AA, therefore there is insufficient information to include a secondary component. By only considering a single power-law through the \lya{} line profile, we may bias our results slightly on fitting the \lya{} broad component. However, provided we are consistent in our usage of this QSO continuum between this fitting approach and the reconstruction method, this should not impact our results.

As mentioned in the previous section, we model each emission line component with a Gaussian profile. Following \citet{Kramer:2009p920} the total flux for each component can be defined as,
\begin{eqnarray} \label{eq:flux}
F_{i} = a_{i}\,{\rm exp}\left[ - \frac{(\lambda - \mu_{i})^{2}}{2\sigma^{2}_{i}}\right],
\end{eqnarray}
where $a_{i}$ describes the amplitude of the line peak, $\mu_{i}$ is the location of the line centre in \AA, $\sigma_{i}$ is the width of the line in \AA~and the subscript `$i$' denotes the specific line species (e.g. \lya{}). Note that within this work, the peak amplitude is always normalised by the continuum flux at $1450$\AA, $f_{1450}$, therefore it is always a dimensionless quantity. More intuitively, the line centre location, $\mu_{i}$, can be written in terms of a velocity offset relative to the systemic line centre,
\begin{eqnarray}
v_{{\rm shift},i} = c\,\frac{\mu_{i} - \lambda_{i}}{\lambda_{i}} \,{\rm km/s},
\end{eqnarray}
and the line width can be expressed as
\begin{eqnarray} \label{eq:width}
\sigma_{i} = \lambda_{i}\left(\frac{{\rm LW}}{c}\right)\,{\rm \AA},
\end{eqnarray}
where LW is the line width measured in km/s. Throughout Equations~\ref{eq:flux}-\ref{eq:width}, both $\lambda$ and $\lambda_{i}$ are measured in the rest frame. 

Each Gaussian line component can therefore be fully described by its three component parameters, the line width, peak amplitude and velocity offset. In total, we fit each QSO with two continuum parameters, two double component Gaussians (\lya{} and \civ{}), and 9 single component Gaussian profiles, resulting in a total of 41 continuum and emission line parameters.

\subsection{Identifying absorption features} \label{sec:absorb}

Intervening diffuse neutral hydrogen or larger column density absorbers (smaller than DLAs) along the line of sight can produce narrow absorption features. Furthermore, metal pollution in stronger absorption systems can additionally result in narrow absorption features appearing in the observed profiles of the emission lines.

These narrow features, if not measured and accounted for, can artificially bias the shape and peak amplitude of the emission line components. Therefore, in this section we outline our approach for identifying these features in a clean and automated manner:
\begin{itemize}
\item Identify all flux pixels that are a local minima within a 2\AA~region surrounding the central pixel in question. This choice of 2\AA~is arbitrary, but is selected to be sufficiently broad to ignore features that might arise from noise fluctuations.
\item Construct a horizontal line of constant flux that begins from the global flux minimum.
\item Incrementally increase this line of constant flux, recording the depth and width of each absorption feature enclosed by the line of constant flux.
\item If the depth becomes larger than 3$\sigma$ (5$\sigma$ in the vicinity of \lya{}) of the observed error in the flux at that pixel and the candidate absorption line has remained isolated (i.e. not overlapped with a nearby feature) it is then classified as an absorption line.
\end{itemize}

Our adopted choice of 3$\sigma$ (5$\sigma$ near \lya{}) arises after rigorously testing this pipeline against the absorption features that were visually identified, until the vast majority of the absorption lines could be robustly located by this procedure. Across the full QSO sample, we found a broad range in the number of features identified within each individual QSO, from only a few up to $\sim40$. For each identified absorption feature, a single Gaussian (described by its own three parameters) is assigned and is simultaneously fit with the continuum and emission line profiles outlined previously. Therefore, the total number of parameters to be fit per QSO varies.

\subsection{MCMC sampling the QSO template} \label{sec:MCMC}

We fit each QSO within a Bayesian MCMC framework, using the $\chi^{2}$ likelihood function to determine the maximum likelihood fit to the QSO spectrum. This choice enables us to fully characterise any potential model degeneracies between our model parameters, while also providing the individual probability distribution functions (PDFs) for each model parameter. In this work, we utilise the publicly available MCMC python code \textsc{CosmoHammer} \citep{Akeret:2012p842} built upon \textsc{EMCEE} \citep{ForemanMackey:2013p823} which is based on the affine invariant MCMC sampler \citep{Goodman:2010p843}.

\begin{figure*} 
	\begin{center}
		\includegraphics[trim = 0.25cm 0.75cm 0cm 0.7cm, scale = 0.495]{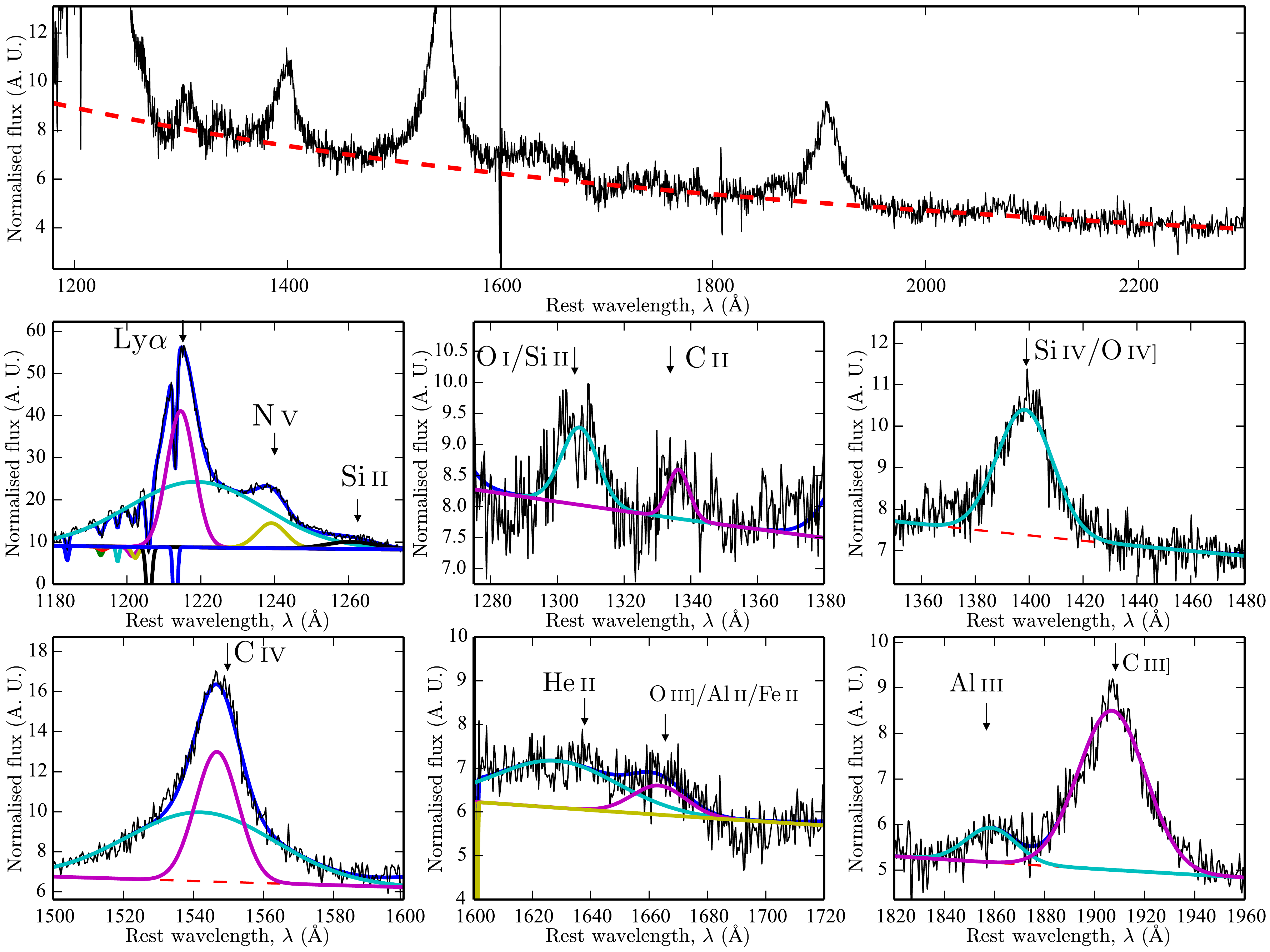}
	\end{center}
\caption[]{An example, zoomed in figure of the MCMC QSO template fitting of a BOSS (SDSS-III) spectrum (`spec-4185-55469-0076' a $z=2.478$ QSO). For all spectra we translate the flux into arbitrary units (1 A. U. = $10^{-17}\,{\rm erg\,cm^{-2}\,s^{-1}\,\AA^{-1}}$). \textit{Top:} The full QSO spectrum used within our fitting procedure as a function of the rest frame wavelength, $\lambda$. The red dashed curve corresponds to the two parameter continuum (see Equation~\ref{eq:continuum}). \textit{Middle left:} The two component fit to the \lya{} emission profile highlighted by the broad (cyan) and narrow (magenta) components and a single Gaussian profiles for N\,{\scriptsize V} (yellow) and Si\,{\scriptsize II} (black). A series of `absorption' features were identified and fit to the spectrum (shown below the continuum level). \textit{Middle centre}: Two low ionisation lines, O\,{\scriptsize I} (cyan) and C\,{\scriptsize II} (magenta). \textit{Middle right:} Single component fit to the \sioiv{} doublet. \textit{Bottom left:} Double component fit to \civ{} highlighted by the broad (cyan) and narrow (magenta) components. \textit{Bottom centre:} Low ionisation lines, He\,{\scriptsize II} (cyan) and O\,{\scriptsize III} (magenta). \textit{Bottom right:} Single component fit to \ciii{} (magenta) and single Gaussian Al\,{\scriptsize III} component (cyan). The maximum likelihood fit to this spectrum was $\chi^{2} = 3110$, for which we recovered a $\chi_{\rm red}^{2} = 1.1$ (with 2828 degrees of freedom, 2899 bins and 71 free parameters).}
\label{fig:QSOexample}
\end{figure*}

Assuming flat priors across our $>50$ model parameters and fitting the full high-resolution BOSS spectrum within the wavelength range ($1180{\rm \AA} < \lambda < 2300{\rm \AA}$) simultaneously is computationally inefficient. Instead, we perform an iterative procedure to boost the computational efficiency which we outline below:
\begin{itemize}
\item Normalise the full QSO spectrum at 1450\AA~and then fit the two continuum parameters, $f_{1450}$ and $\alpha_{\lambda}$ within a set of selected wavelength ranges of the full spectrum which are minimally contaminated by emission lines. We choose to fit the QSO continuum within the regions [1275, 1295], [1315, 1330], [1351, 1362], [1452, 1520], [1680, 1735], [1786, 1834], [1970, 2040] and [2148, 2243]~\AA.
\item Break the full QSO spectrum up into wavelength regions centred around the emission lines. We choose [1180, 1350]~\AA~centred on \lya{}, [1350, 1450]~\AA~centred on \sioiv{}, [1450, 1700]~\AA~centred on \civ{} and [1700, 1960]~\AA~centred on \ciii{}.
\item Within each of these four regions, we take the continuum estimated from above, and then fit all emission lines and any absorption features which fall within the respective wavelength ranges. For each of these regions we then perform an MCMC fit.
\item From the individual PDFs for each parameter we construct a flat prior across a much narrower allowed range, driven by the width of the individual distributions from the fitting above.
\item Finally, we fit the entire QSO spectrum using the entire model parameter set and recover a maximum likelihood model which describes the full spectrum.
\end{itemize}

In Figure~\ref{fig:QSOexample}, we provide an example of one of the BOSS QSOs from our full sample. In this figure we provide zoomed in panels of both the QSO continuum (red dashed curve, top panel) and the various emission lines which are simultaneously fit within our MCMC approach across all other panels. This figure highlights that any notable absorption features have been identified by our pipeline and accurately characterised (e.g. the \lya{} line profile in the left panel of the middle row). From Figure~\ref{fig:QSOexample} we note that this approach is able to fit the full spectrum. The maximum likelihood we obtained was $\chi^{2} = 3110$, for which we had 2899 bins (from the raw spectrum) and 71 free parameters, corresponding to a reduced $\chi^{2}$ of 1.1. For reference, this took $\sim$1 hour on a single processing core, which can be rapidly improved if a binned spectrum is used.

\subsection{QSO fitting quality assessment} \label{sec:QA}

After fitting the entire sample with our full MCMC QSO fitting pipeline, we are in a position to construct a higher fidelity sample of QSOs for constructing our covariance matrix and investigating the correlations amongst the emission line parameters. This is performed by visually inspecting our QSO sample, applying a simple selection criteria. Following the completion of this selection process, we produce two separate QSO samples, one classified as `good' and another classified as `conservative', the details of which we discuss below (in Appendix~\ref{sec:QSO_QA} we provide a few select examples to visually highlight this subjective process). The criteria are outlined as follows:
\begin{itemize}
\item We remove all QSOs with a poor characterisation of the continuum. These include QSOs with a positive spectral index, or a clear departure from a single power-law continuum. Only a handful of QSOs exhibit this behaviour. See Figure~\ref{fig:QA_RemovedContinuum} for some examples.
\item We further remove any QSOs which have either (i) missing sections of flux that overlap with any of the emission lines (ii) a sufficient number of absorption features which cause a loss of confidence in the fitting of the \lya{} profile (iii) either an intervening dense neutral absorber or sufficiently strong/broad absorption blueward of line centre which can impact the broad component of either the \lya{} or the \civ{} lines (iv) a \lya{} line region that is not well fit or characterised by our double component Gaussian, which might arise from numerous absorption profiles or the lack of a prominent \lya{} peak.
\item Following the removal of these contaminants from our sample, we are left with 2,653 QSOs, which we call our `conservative' sample\footnote{Our naming convention of 'good' and 'conservative' refers to the tightness of constraints recovered from the reconstruction procedure. The 'conservative' sample, with less assumptions on data quality, produces slightly broader errors relative to the 'good' sample.}. These can still include QSOs which might have absorption features centred on the \lya{} line centre, or absorption line complexes which might contaminate large sections of the \lya{} line profile (see Figure~\ref{fig:QA_Conservative} for examples).
\item We then apply a secondary criteria, which preferentially selects QSOs which contain (i) fewer absorption features and (ii) no absorption features on the line centre. This results in a final sample of 1673 QSOs, which we refer to as our `good' sample (see Figure~\ref{fig:QA_Good} for examples).
\end{itemize} 

The major difference between the `good' and `conservative' sample is that the `conservative' sample will contain QSO spectra for which we have less confidence in either the identification and fitting of one, or several of the emission lines due to the presence of absorption features. Ultimately, if our claim that any correlations amongst the emission lines are a universal property of the QSOs then the covariance matrices of the two should be almost identical, with the `conservative' sample containing additional scatter (slightly weaker correlations). In the next section, we construct our covariance matrices and investigate this further.

\section{Data Sample Covariance} \label{sec:Covariance}

With the refined, quality assessed QSO spectra from the previous section, we now construct our covariance matrix to characterise all correlations amongst the various emission lines. 

\subsection{The covariance matrix} \label{sec:covariance}

In performing the quality assessment of our QSO spectra, we note that a number of the weaker emission lines are not always well characterised or resolved. This becomes more prevalent for the QSOs nearer to our sample limit of S/N~$=15$, which correspond to the highest density of QSOs in our sample. It would be interesting to investigate correlations between the strong high ionisation lines and the weaker low ionisation lines, as well as correlations amongst these two classifications. However, since these lines are not always readily available in our QSOs we refrain from doing so. Regardless, by still attempting to fit these weaker lines we retain the flexibility of the MCMC approach, and more importantly this enables the QSO continuum to be estimated to a higher accuracy. 

\begin{figure*} 
	\begin{center}
		\includegraphics[trim = 0.15cm 0.3cm 0cm 0.5cm, scale = 0.94]{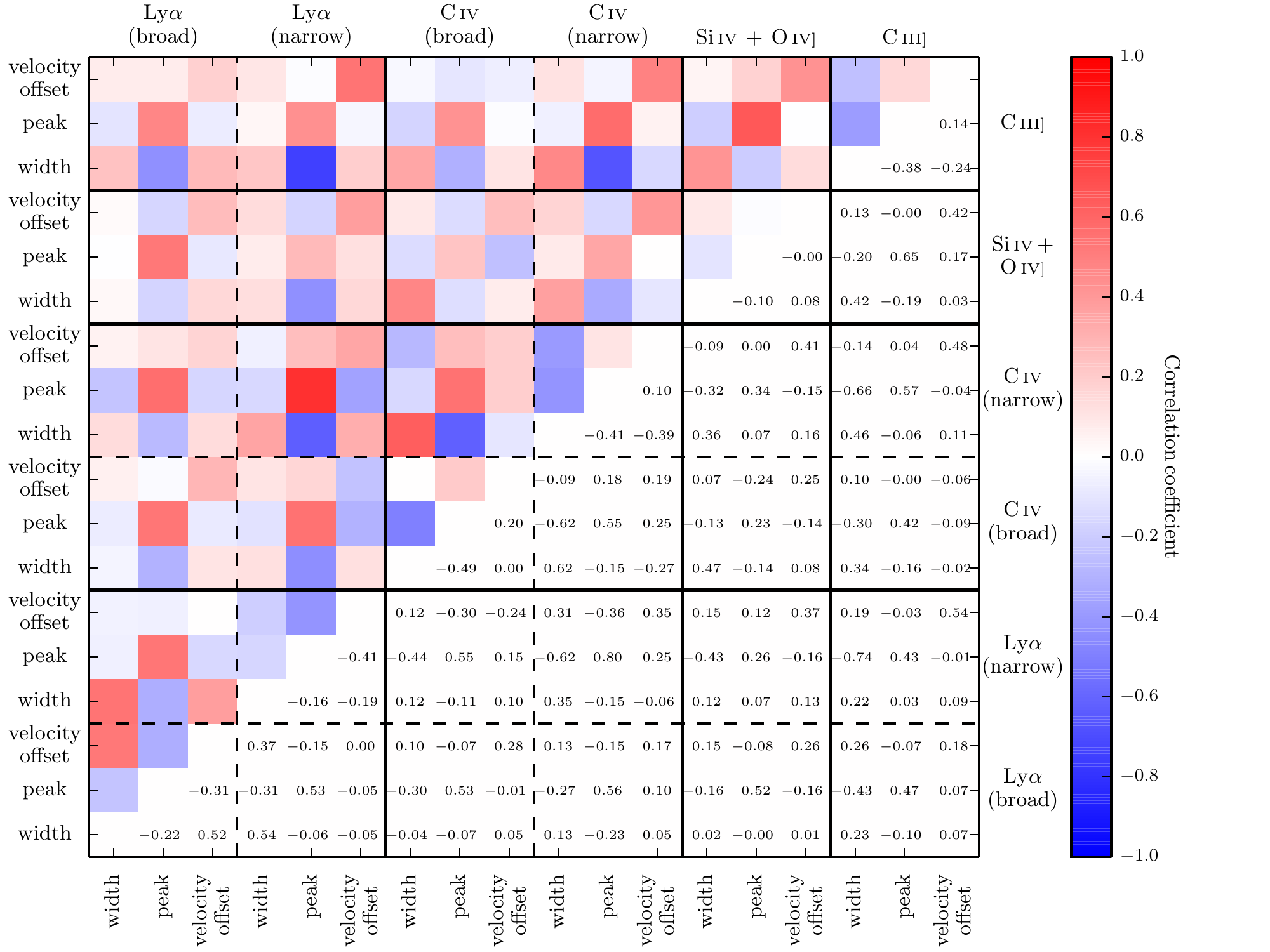}
	\end{center}
\caption[]{The correlation coefficient matrix (correlation coefficients listed in the lower half) constructed from the `good` sample of 1673 QSOs. This $18\times18$ matrix contains the double component Gaussians of \lya{} and \civ{} and the single component Gaussians for \sioiv{} and \ciii{}. Each Gaussian component contains three parameters, its peak width, peak height and velocity offset from systemic (\lya{} + \civ{} + \sioiv{} + \ciii{} = ($2\times3$) + ($2\times3$) + 3 + 3 = 18 parameters).}
\label{fig:CovarianceMatrix}
\end{figure*}

In light of this, we construct our covariance matrix from a subset of these emission lines: the most prominent emission lines that should always be resolvable in a lower S/N or lower resolution QSO spectrum (as in the near-IR for $z>6$ QSOs). The lines we identify are \lya{}, \sioiv{}, \civ{} and \ciii{}. Since for \lya{} and \civ{} we allow a broad and narrow component, we recover an 18$\times$18 covariance matrix. We choose to exclude the \nv{} line from our covariance matrix for two reasons: (i) a clear, identifiable \nv{} line is not always present in our QSO spectra, therefore including it would artificially reduce any visible correlation and (ii) the ultimate goal of this work is the reconstruction of the \lya{} line whereby including the \nv{} line to the reconstruction process would only increases the complexity (the \nv{} line should in principle be recoverable from the high-$z$ or \lya{} obscured QSO).

In Figure~\ref{fig:CovarianceMatrix}, we present the correlation coefficient matrix, which is obtained from the full covariance matrix from our `good' sample of 1673 QSOs. In constructing this, we assume the standard format for the covariance matrix, $\bmath{\Sigma}_{ij}$, given by,
\begin{eqnarray}
\bmath{\Sigma}_{ij} = \frac{1}{N-1}\sum^{N}_{i}(\textbfss{X}_{i} - \bmath{\mu}_{i})(\textbfss{X}_{j} - \bmath{\mu}_{j}),
\end{eqnarray}
where $\textbfss{X}_{i}$ is the data vector for the full QSO sample and $\bmath{\mu}$ is the mean data vector for the $i$th parameter. For this data vector we use the values from the parameter set which provide the maximum likelihood fit to each QSO. The correlation coefficient matrix, $\textbfss{R}_{ij}$, is then defined in the standard way,
\begin{eqnarray}
\textbfss{R}_{ij} = \frac{\textbfss{C}_{ij}}{\sqrt{\textbfss{C}_{ii}\textbfss{C}_{jj}}},
\end{eqnarray}
where each diagonal entry, $\textbfss{R}_{ij}$, corresponds to the correlation coefficient between the $i$th and $j$th model parameters.

Note, for the covariance matrix, we do not include the two continuum parameters. Firstly, for the parameters defining the emission line peak we define this parameter as the normalised peak value, where it is normalised by the continuum flux at $f_{1450}$. Therefore, if included, this would be completely degenerate with the peak amplitudes. Some correlation with continuum spectral index is expected, given that external (to the broad-line region) reddening will simultaneously weaken the bluer UV lines and redden the continuum spectral index. However, mild reddening is seen in only $\sim10-20$~per cent of SDSS quasars \citep[e.g.][]{Richards:2003p4639,Hopkins:2004p4640} so this has little effect on the correlations (we find very weak correlations on the order of 10 per cent). Therefore we do not report them, as these provide no additional information with respect to the individual line correlations.

\subsection{Interpreting the covariance matrix} \label{sec:interpretation}

In order to aid the interpretation of the correlation matrix, we divide Figure~\ref{fig:CovarianceMatrix} into the four emission line species (denoted by black curves), while the narrow and broad line species are further separated by black dashed lines. Positive correlations are represented by a decreasing (weakening) shading of red, with white representing no correlation and an increasing blue scaling denotes a strengthening of the anti-correlation. Using the upper half of the correlation matrix enables a faster analysis of the correlation patterns amongst the emission line parameters, and the lower half reports the numerical value of the correlation (or anti correlation). For the most part, each 3$\times$3 sub-matrix returns the same correlations and anti-correlations amongst the peaks, widths and velocity offsets, but with varying degrees of strength.

\begin{figure} 
	\begin{center}
		\includegraphics[trim = 0.5cm 0.7cm 0cm 0cm, scale = 0.58]{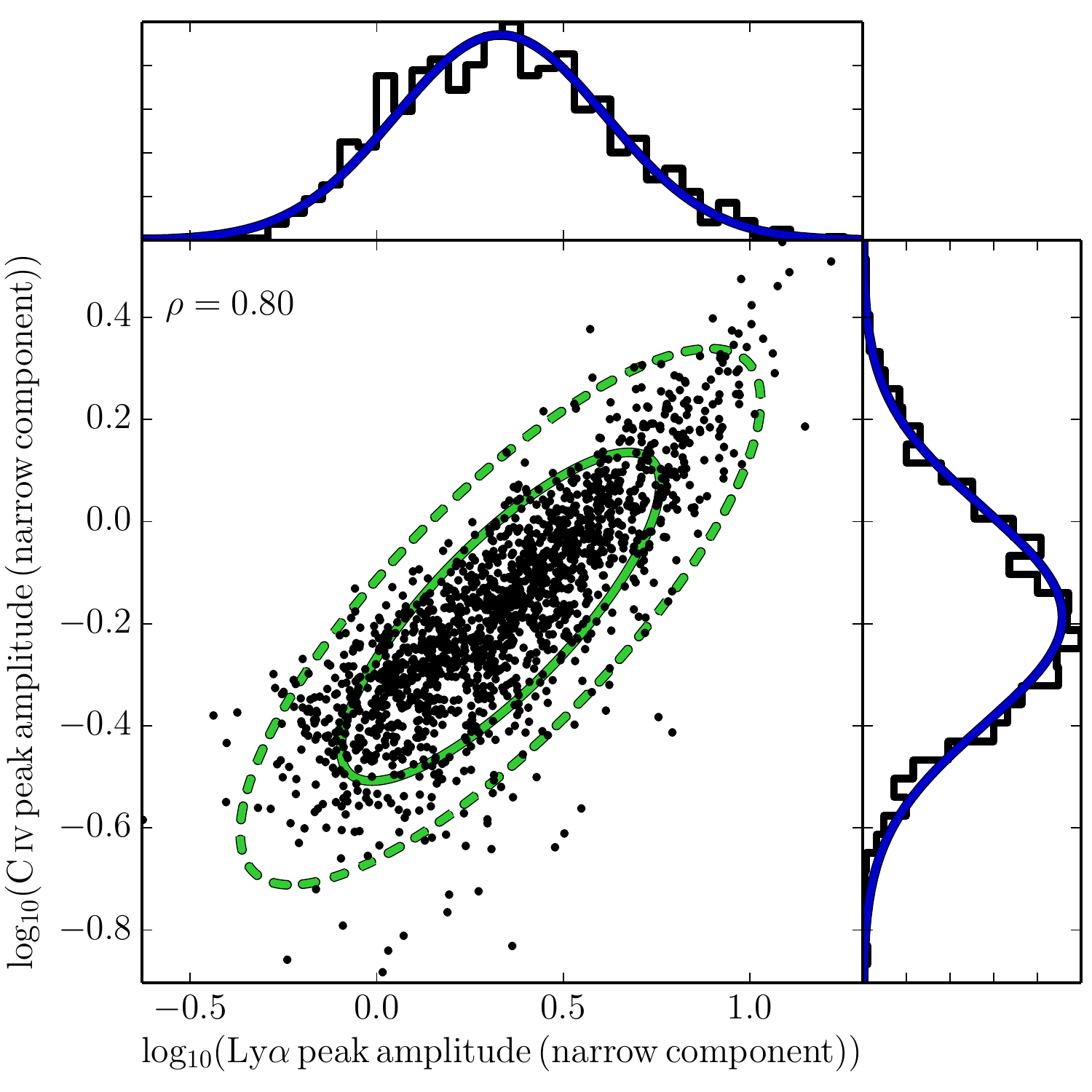}
	\end{center}
\caption[]{A 2D scatter plot of the correlation between the peak amplitudes of the narrow components of the \lya{} and \civ{} emission lines for our `good' sample of 1673 QSOs. We recover a correlation coefficient of $\rho=0.80$, indicative of a strong, positive correlation. The green solid and dashed curves correspond to the 68 and 95 per cent 2D marginalised joint likelihood contours, respectively. The histograms (black curves) correspond to the recovered PDFs for each of the two parameters, while we also provide Gaussian curves (blue) representative of the scatter in the fit parameters (not a direct fit). Note that for the peak amplitude, we normalise by $f_{1450}$, therefore these are dimensionless.}
\label{fig:PeakCorrelation}
\end{figure}

Firstly, we note that with this correlation matrix it is straightforward to notice that all pairs of peak height parameters are positively correlated. This is to be expected as the peak amplitude is positively correlated with the QSO luminosity. This correlation between the peak heights is the strongest of the trends. Most importantly for this work, the correlations amongst the \lya{} peak height parameters are the strongest. For example, we find the strongest correlation ($\rho = 0.8$) between the peak height of the \lya{} narrow component and the associated peak height of the \civ{} narrow component. In Figure~\ref{fig:PeakCorrelation}, we provide the 2D scatter plot for this strong correlation (central panel) and the 1D marginalised PDFs for the \lya{} (top) and \civ{} (right). The green solid and dashed contours in the central panel denote the 68 and 95 per cent 2D marginalised joint likelihood contours, which describe the relative scatter amongst our sample of QSOs. In the 1D marginalised PDFs, the solid black curves are the histograms of the sample of QSOs, while the blue solid curves are an approximated 1D Gaussian for that associated parameter. Note in this figure, and in subsequent figures, these blue curves are not a fit directly to the raw data, rather instead they are approximations of a Gaussian PDF with 1$\sigma$ scatter equivalent to the raw data.

\begin{figure} 
	\begin{center}
		\includegraphics[trim = 0.5cm 0.6cm 0cm 0cm, scale = 0.59]{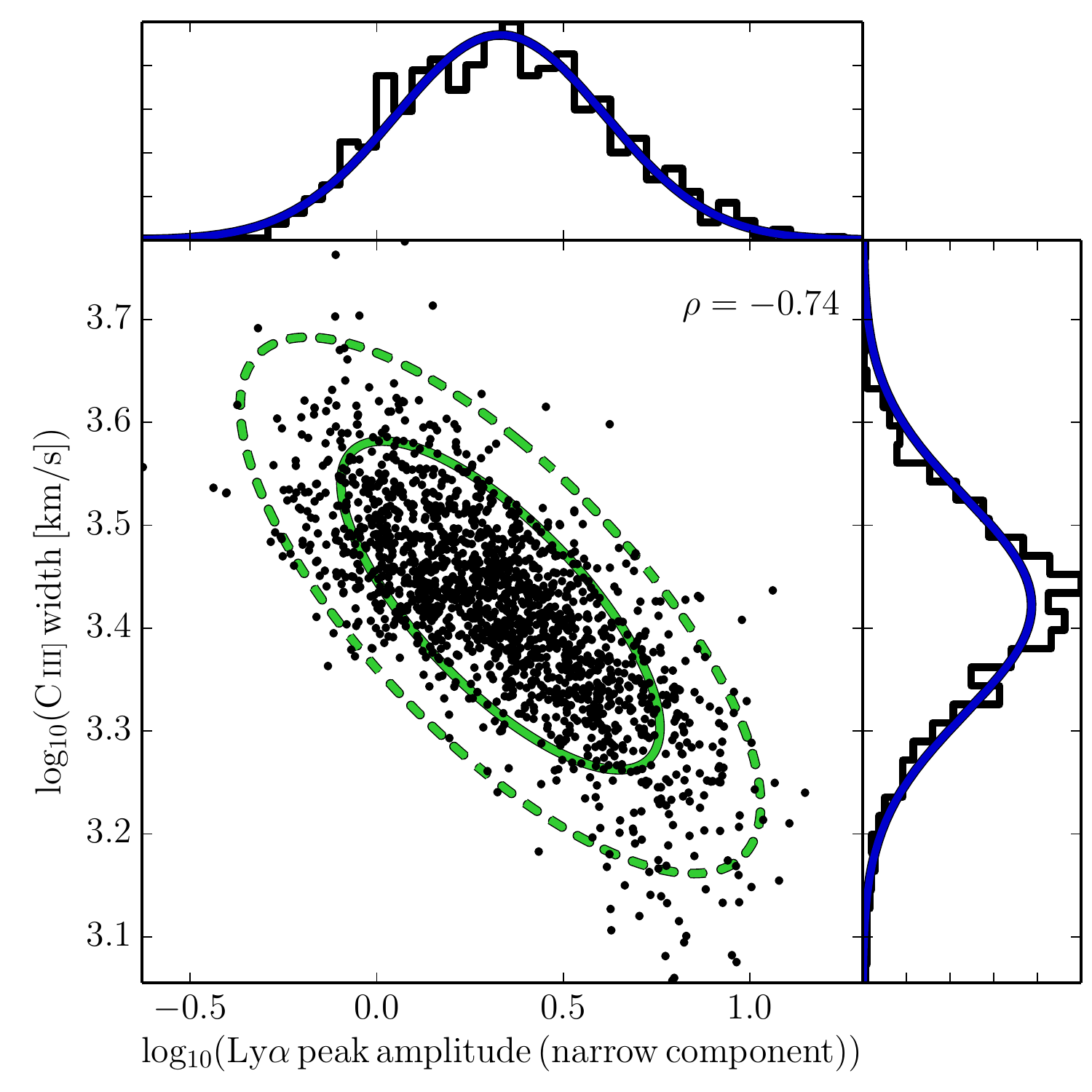}
	\end{center}
\caption[]{A 2D scatter plot of the correlation between the narrow component of the \lya{} peak amplitude and line width of the \ciii{} emission line for our `good' sample of 1673 QSOs. We recover a correlation coefficient of $\rho=-0.74$, indicative of a strong anti-correlation. Histograms, blue curves and the green contours are as described in Figure~\ref{fig:PeakCorrelation}. As in Figure~\ref{fig:PeakCorrelation}, the peak amplitude is normalised by $f_{1450}$ and is therefore dimensionless.}
\label{fig:PeakWidthCorrelation}
\end{figure}

Returning to the correlation matrix, we additionally recover a relatively strong trend (anti-correlation) between the peak height and the line width. In Figure~\ref{fig:PeakWidthCorrelation}, we provide the strongest of these anti-correlations ($\rho=-0.74$) which is between the \lya{} peak amplitude of the narrow line component and the width of the single component \ciii{} line. We observe, that for an increasing peak amplitude for the \lya{} narrow component, a decreasing of the \ciii{} line width. This behaviour could be inferred as a result of the Baldwin effect \citep{Baldwin:1977p5910}. For an increasing peak amplitude (i.e. QSO luminosity), we expect a weaker broad-line emission. Physically, a plausible scenario to describe this could be the over-ionisation of the inner broad-line region by the continuum in high luminosity QSOs, resulting in carbon being ionised into higher order species (i.e.\ no \civ{}/\ciii{}). For these high luminosity QSOs, the high ionisation lines may then predominantly arise at larger radii where the velocity dispersion is lower, producing smaller line widths \cite[e.g.][]{Richards:2011p6031}. Alternately, it could arise from systematics from our line fitting. As a result of the decreasing equivalent widths with increasing QSO luminosity, it could be that the single component Gaussian for the \ciii{} does not characterise the line profile as accurately. With a less prominent broad-line component, the emission from the wings could be underestimated, producing a narrower \ciii{} line.

We additionally recover a trend for a positive correlation between the widths of the various line species, though this is a relatively weaker trend. However, for the broad and narrow components for the same line species, we find more moderate correlations ($\rho=0.54$ for \lya{} and $\rho=0.62$ for \civ{}). It is difficult to interpret these correlations though, as these broad and narrow lines are simultaneously fit, and therefore in principle could be degenerate. 

\citet{Shang:2007p4862} investigated correlations amongst the \lya{}, \civ{} and \ciii{} lines for a significantly smaller sample of 22 QSOs. These authors recover moderate to strong correlations between a few of their emission line width full width half maxima (FWHMs). They find \lya{}--\civ{} ($\rho=0.81$) and \civ{}--\ciii{} ($\rho=0.53$). If we equate this FWHM as the width of the narrow component of our emission line profiles we find correlations between the same species of \lya{}--\civ{} ($\rho=0.35$) and \civ{}--\ciii{} ($\rho=0.46$)\footnote{By converting our fitted emission line parameters (both broad and narrow) into a line profile FWHM width, we find correlations similar in strength to those recovered purely from the narrow line component.}. While we recover an equivalent correlation for the \civ{}--\ciii{} lines, we find a significant discrepancy for the \lya{}--\civ{} lines. This potentially can be explained as an artificially large correlation owing to their small number of objects (22 compared to our 1683 QSOs)\footnote{Note that, if we adopt a systemic redshift measured purely from a single ionisation line (such as the \mgii{} redshift) rather than the pipeline redshift used throughout this work, we can recover a significantly stronger correlation of \lya{}--\civ{} ($\rho=0.7$).}. Supporting this hypothesis, in \citet{Corbin:1996p5640} for a larger sample of 44 QSOs, this is reduced to $\rho=0.68$ for \lya{}--\civ{}.

\begin{figure} 
	\begin{center}
		\includegraphics[trim = 0.5cm 0.6cm 0cm 0cm, scale = 0.58]{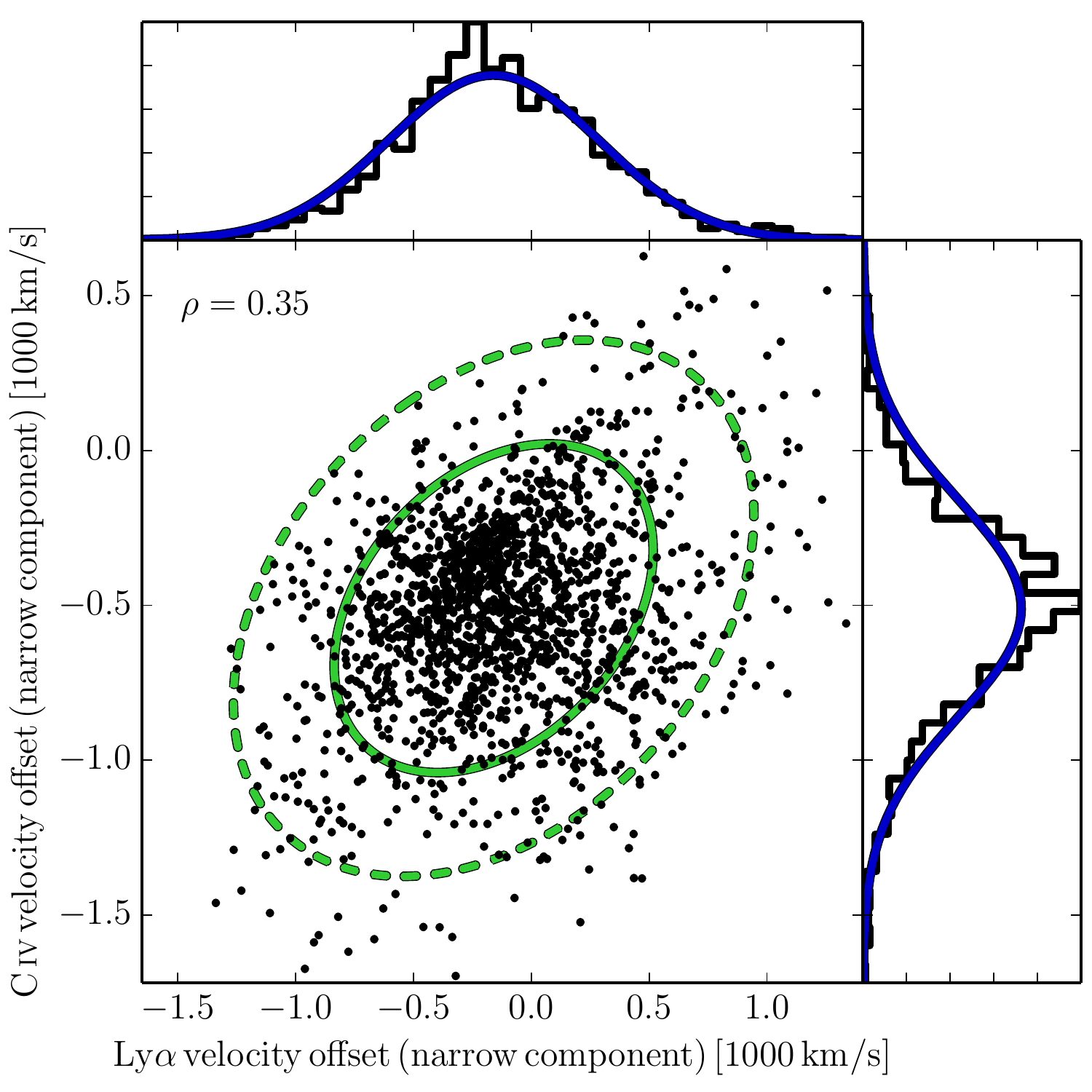}
	\end{center}
\caption[]{A 2D scatter plot of the correlation between the velocity offsets of the \lya{} and \civ{} emission lines for our `good' sample of 1673 QSOs. We recover a correlation coefficient of $\rho=0.35$, indicative of a moderate correlation, which is counter to the much stronger correlation hinted at from the smaller QSO samples of \citet{Shang:2007p4862} and \citet{Kramer:2009p920}. Histograms, blue curves and the green contours are as described in Figure~\ref{fig:PeakCorrelation}.}
\label{fig:VelocityCorrelation}
\end{figure}

Additionally, \citet{Kramer:2009p920} recovered a moderate positive correlation ($\rho=0.68$) between the velocity offsets for the \lya{}--\civ{} components. For the same combination, \citet{Shang:2007p4862} found a stronger correlation of $\rho=0.81$. In Figure~\ref{fig:VelocityCorrelation}, we provide our weaker results between the narrow components of the \lya{}--\civ{} lines for which we recover a moderate correlation of $\rho=0.35$. Once again, this lack of an equivalently strong correlation in our sample could arise purely from the fact that our sample contains two orders of magnitude more QSOs. \citet{Shang:2007p4862} equivalently quote correlations of $\rho=0.45$ (\lya{}--\ciii{}) and $\rho=0.39$ (\civ{}--\ciii{}). Using the narrow lines for \lya{} and \civ{}, we find similarly moderate, but slightly stronger correlations of $\rho=0.54$ (\lya{}--\ciii{}) and $\rho=0.48$ (\civ{}--\ciii{}). 

In the construction of the covariance matrix, and associated correlation matrix we have only used the maximum likelihood values, and ignored the relative errors for each parameter from the marginalised 1D PDFs. Our reasoning for this is that the amplitude of the error on the individual parameters is small with respect to the scatter that arises for each parameter across the full QSO sample. In principle however, we could perform an MCMC sampling of the full covariance matrix, allowing the means and all correlation coefficients to be free parameters. This would allow a more accurate characterisation of both the individual errors for each parameter and the total scatter across the full QSO sample, potentially tightening the correlations amongst the emission line parameters. However, this would require fitting $\frac{1}{2}N(N+1)$ parameters encompassing both the means of the individual parameters and all correlation coefficients (in our case, 171 free parameters).

\subsection{Correlation of the QSO continuum parameters}

\begin{figure} 
	\begin{center}
		\includegraphics[trim = 0.5cm 0.7cm 0cm 0cm, scale = 0.58]{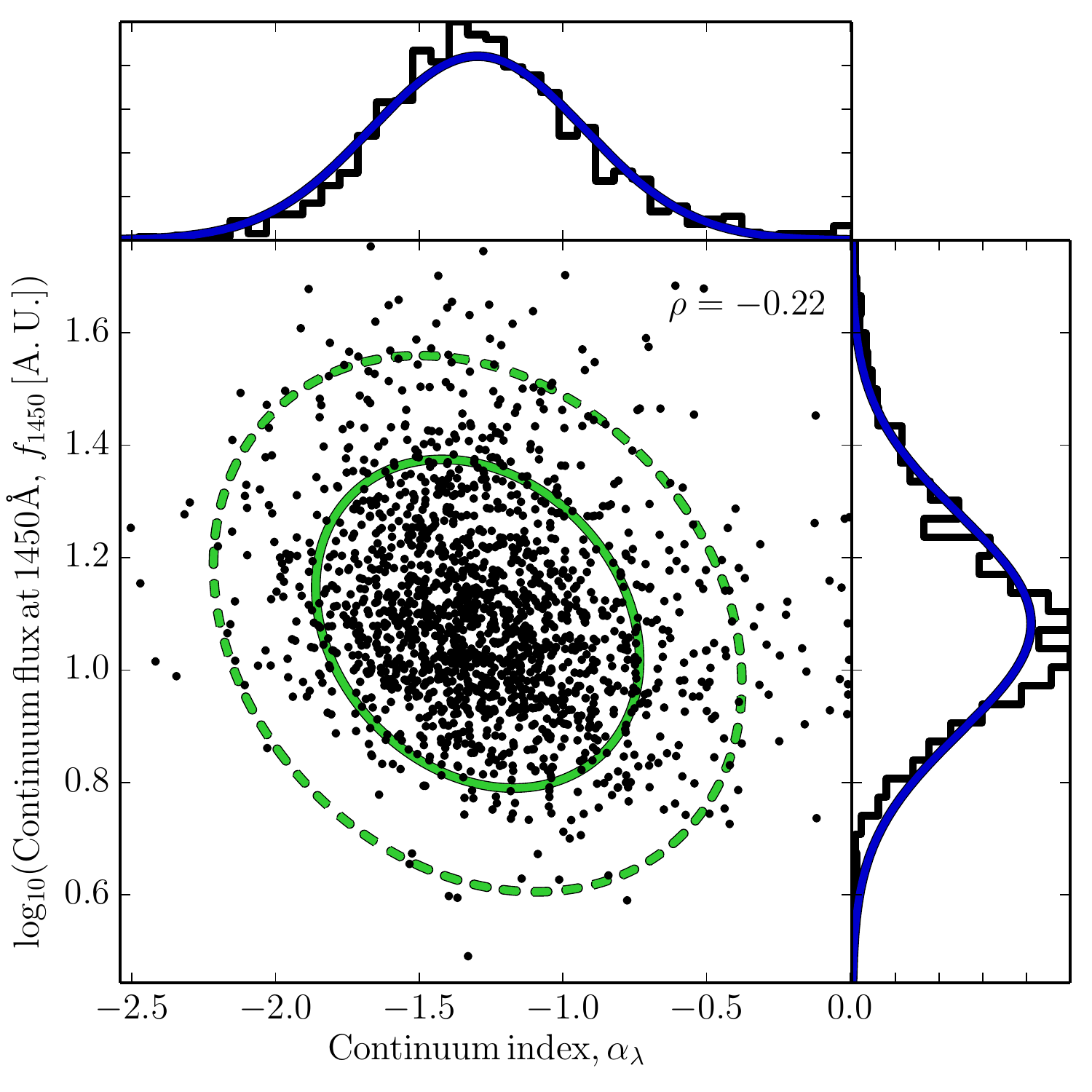}
	\end{center}
\caption[]{A 2D scatter plot highlighting the correlation between the two QSO continuum parameters: the continuum spectral index, $\alpha_{\lambda}$ and the continuum flux at 1450\AA, $f_{1450}$ (for which we define 1 A. U. = $10^{-17}\,{\rm erg\,cm^{-2}\,s^{-1}\,\AA^{-1}}$), for our sample of 1673 QSOs (the `good' sample). We recover a correlation coefficient of $\rho=-0.22$, indicative of a weak anti-correlation. Histograms, blue curves and the green contours are as described in Figure~\ref{fig:PeakCorrelation}.}
\label{fig:Continuum}
\end{figure}

Though the continuum parameters are not included in the covariance matrix in Figure~\ref{fig:CovarianceMatrix}, in Figure~\ref{fig:Continuum} we provide the 2D scatter between these two parameters. In our `good' sample of QSOs, we recover only a relatively weak anti-correlation ($\rho=-0.22$) between the QSO continuum spectral index, $\alpha_{\lambda}$ and the normalisation at 1450~\AA~($f_{1450}$). This weak anti-correlation likely arises due to dust reddening, which causes a drop in the normalisation, $f_{1450}$ and shallower spectral slopes. While both of these QSO parameters are well characterised by a Gaussian, the respective scatter in each parameter is considerable.

For our sample of `good' (`conservative') QSOs, we recover a median QSO spectral index of $\alpha_{\lambda} = -1.30\pm0.37$ ($ -1.28\pm0.38$). In contrast, \citet{Harris:2016p5028} construct a QSO composite spectrum with a wavelength coverage of ($800{\rm \AA} < \lambda < 3300{\rm \AA}$) from the same BOSS DR12 sample. Using $\sim100,000$ QSOs, these authors find a median spectral index for their QSO sample of $\alpha_{\lambda} = -1.46$, consistent with our results within the large scatter. However, these authors only fit the QSO continuum between 1440-1480\AA~and 2160-2230\AA. Not fitting to the same spectral region likely results in a different recovered spectral slope. Furthermore, our estimate of $\alpha_{\lambda} = -1.30$ ($\alpha_{\nu} = -0.70$) is also consistent with the lower redshift samples of \citet{Scott:2004p5709} ($\alpha_{\nu}=-0.56\substack{+0.38 \\ -0.28}$), \citet{Shull:2012p5716} ($\alpha_{\nu}=-0.68\pm0.14$) and \citet{Stevans:2014p5726} ($\alpha_{\nu}=-0.83\pm0.09$). Note, within all these works, considerable scatter in the spectral index is also prevalent.

\subsection{Potential sample bias} \label{sec:Bias}

In Sections~\ref{sec:covariance} and~\ref{sec:interpretation}, we presented the correlation matrix and discussions on the relative trends between the emission lines and their relative strengths. However, these results were drawn from our refined, quality assessed `good' QSO sample. In order to guard against a potential bias which may have arisen following our specific selection process, we additionally construct the correlation matrix for our `conservative' QSO sample. This `conservative' sample contains $\sim1000$ additional QSOs which are defined to be less robust than required for our `good' sample. Therefore, this `conservative' sample should contain more scatter amongst the recovered parameters, which would notably degrade the strength of the correlations relative to the `good' sample if we have artificially biased our results.

\begin{figure*} 
	\begin{center}
		\includegraphics[trim = 0.15cm 0.3cm 0cm 0.5cm, scale = 0.94]{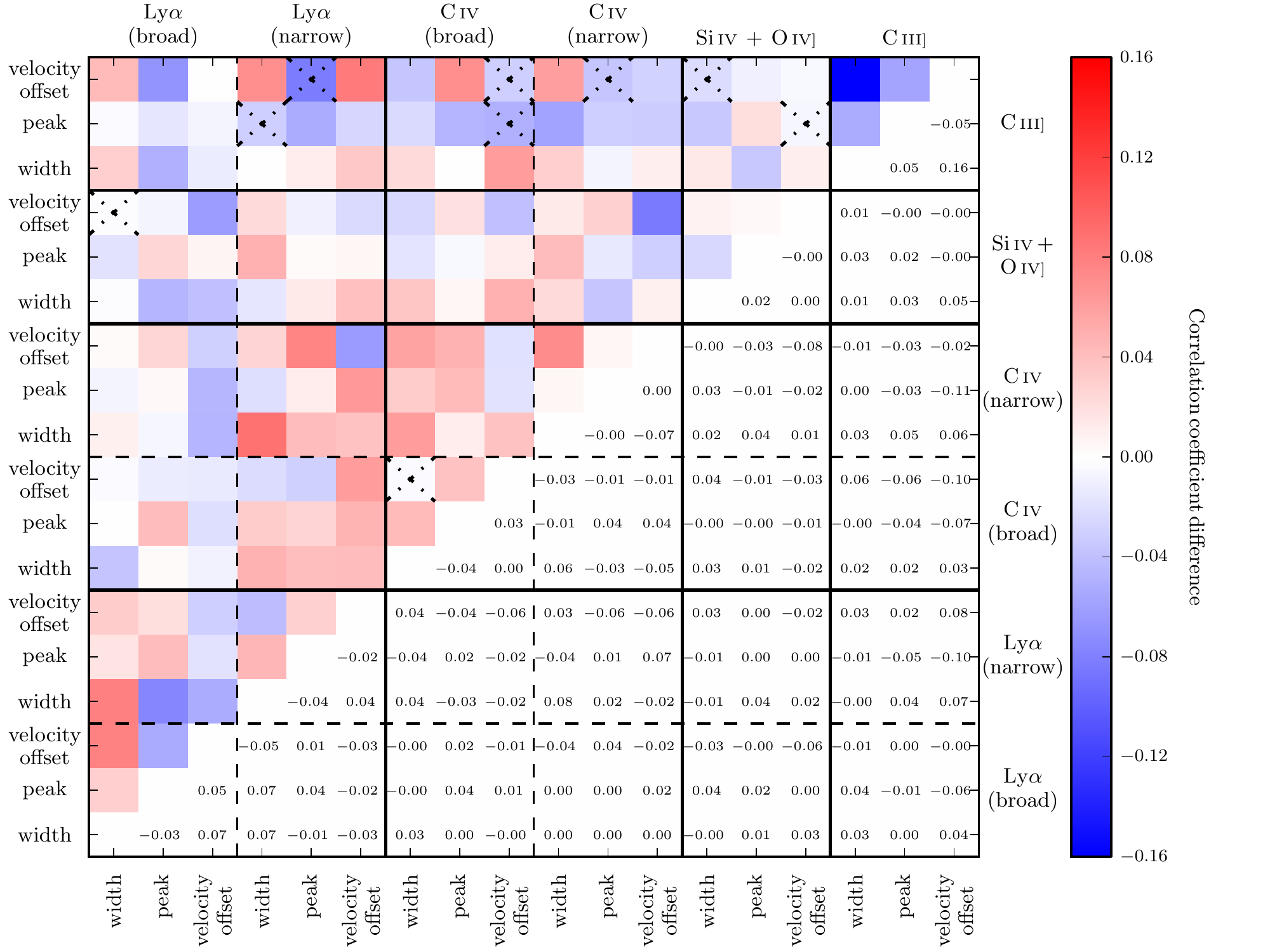}
	\end{center}
\caption[]{The amplitude of the difference of the correlation coefficients for the $18\times18$ covariance matrix between the `good' (1673) and `conservative' (2653) QSO datasets. Positive (red) differences indicate where the correlation in the `good' sample is greater than the `conservative' and negative (blue) indicates when the `conservative' sample is strongest. Note this does not differentiate between a positive or anti-correlation nor the strength of the original correlation, just a strengthening/weakening of the respective correlation. Dot-dashed squares indicate a swap between a positive and negative correlation (typically indicative of the correlation coefficients in either QSO samples being close to zero). The colour bar has been renormalised relative to Figure~\ref{fig:CovarianceMatrix} to indicate the strength of the difference in amplitude.}
\label{fig:CovMatDiff}
\end{figure*}

In Figure~\ref{fig:CovMatDiff} we provide the matrix of the relative difference in the correlation coefficients between the `good' and `conservative' QSO samples. Here, we show the amplitude of the change in correlation coefficient between the two QSO samples. A positive (red) difference is indicative of the `good' sample having a stronger correlation (either positive or anti-correlation), while a negative (blue) difference is indicative of the `conservative' sample having a stronger correlation. Note, squares marked with a dot-dashed cross indicate a change between a positive and anti-correlation, which arises when the correlations are close to zero in either sample.

For the most part, the relative change in the correlation is minor, of the order of $|\Delta\rho| < 0.04$. More importantly, for any of the strong correlations and notable trends we discussed in the previous section, we observe no sizeable differences, with the `good' sample providing on average slightly stronger correlations (as indicated by the prevalence of red squares). For example, the \lya{}--\civ{} peak heights of the narrow component from our `good' sample was $\rho=0.80$, whereas for the `conservative sample it is $\rho=0.79$. Therefore, in the absence of any drastic differences in the strong and notable trends discussed previously between the two samples, it is clear that we have not biased our QSO covariance matrix in the construction of the `good' sample. Given the relative amplitude of these differences between the two QSO samples is small, and tend to be slightly weaker for the `conservative' sample, the reconstructed profile recovered from the covariance matrix should recover effectively the same best-fit \lya{} profile, with slightly broader errors owing to the increased scatter (reduced correlations).

Equally, to be prudent, we considered several other ways to divide our QSO sample. First, we performed the same analysis, comparing instead the `conservative' minus `good' sample (the $\sim1000$ QSOs from the `conservative' sample not classified as `good') and secondly, constructing two equally sized random samples from the `conservative' sample. In both instances, we find the same strong correlations as in Figure~\ref{fig:CovarianceMatrix} with similar amplitude variations between the respective correlation matrices as shown in Figure~\ref{fig:CovMatDiff}. 

Finally, our results on line parameter correlations are sensitive to the choice of redshift estimate and any inherent biases in those estimates. We discuss this issue in detail in Appendix~\ref{sec:systemic_redshift}, but, in summary, we find that while different redshift estimates do result in slightly different covariance matrices, our general conclusions are robust and the redshift estimate we have chosen (the BOSS pipeline redshift, $z_{\rm pipeline}$) performs the best at the \lya{} profile reconstruction (see the next section).

\section{\lya{} Reconstruction} \label{sec:Recon}

We now use this covariance matrix to reconstruct the \lya{} line profile. In this section we outline the reconstruction pipeline, and associated assumptions followed by an example of this approach.

\subsection{Reconstruction method} \label{sec:Reconstruction}

Within our reconstruction approach, we approximate the distribution of the emission line parameters (across the entire data-sample) as a Gaussian\footnote{The choice in adopting a Gaussian covariance matrix is driven by the large computational burden required to perform a full end-to-end Bayesian approach folding in all modelling uncertainties.}. In Figures~\ref{fig:PeakCorrelation}-\ref{fig:VelocityCorrelation}, we observe that for these six parameters shown, this approximation is well motivated. Clearly, showing the 18 individual 1D PDFs for each parameters would be uninformative, however, we confirm by eye that this approximation is valid for all model parameters. Note, that in some of these cases, a Gaussian approximation holds only after taking the logarithm, especially for the normalised peak amplitude (normalised by $f_{1450}$) and the emission line width. Importantly, if anything, these Gaussian approximations have a tendency to slightly overestimate the relative scatter within each parameter (see Figure~\ref{fig:VelocityCorrelation} for example), therefore, this assumption in fact turns out to be a conservative estimate of the true scatter.
The Gaussian nature of the scatter in Figures~\ref{fig:PeakCorrelation}-\ref{fig:VelocityCorrelation} and the conservative overestimation of the errors by our Gaussian approach lends confidence that our approach should not significantly underestimate the  errors that one might recover from a fully Bayesian approach.

In order to perform the reconstruction, we assume that the QSO can be fit following the same procedure outlined in Section~\ref{sec:Fitting}, except now we only fit the QSO red-ward of 1275 \AA. Our choice of $\lambda > 1275$~\AA\,is conservatively selected to be both close enough to \lya{} as possible, but with minimal to no contamination from emission line wings namely \lya{}, N\,{\scriptsize V} or Si\,{\scriptsize II}). Furthermore, given that this approach is best suited for recovering the intrinsic \lya{} profile from a \lya{} obscured or high-$z$ QSO, it is best to be sufficiently far from any possible contamination of the \lya{} line region.

We can then define the $N$ dimensional parameter space (i.e.~our 18 emission line parameters outlined previously) as an $N$ dimensional likelihood distribution given by,
\begin{eqnarray} \label{eq:ML}
\mathcal{L} = \frac{1}{(2\pi)^{N/2}|\bmath{\Sigma}|}{\rm exp}\left[\frac{1}{2}(\bmath{x}-\bmath{\mu})^{\mathsf{T}}\bmath{\Sigma}^{-1}(\bmath{x}-\bmath{\mu})\right].
\end{eqnarray}
Here, $\bmath{\Sigma}$ is the recovered QSO covariance matrix (Section~\ref{sec:covariance}), $\bmath{\mu}$ is the data vector of the means obtained from the full QSO sample for each of the individual line profile parameters and $\bmath{x}$ is the data vector measured from our MCMC fitting algorithm for the individual obscured QSO spectrum. 

After the \lya{} obscured or high-$z$ QSO has been fit following our fitting procedure, the recovered best-fit values for the unobscured emission line parameters are folded into Equation~\ref{eq:ML}. That is, we recover the best-fit estimates of the \sioiv{}, \civ{} and \ciii{} emission lines and evaluate Equation~\ref{eq:ML} to collapse the 18 dimensional likelihood function into a simple, six dimensional likelihood function describing the six unknown \lya{} emission line parameters (two Gaussian components each defined by three parameters). The maximum likelihood of this six dimensional function then describes the best-fit reconstructed profile for the \lya{} emission line, while the full six dimensional matrix contains the correlated uncertainty.

\subsection{Reconstruction example} \label{sec:ReconExample}

\begin{figure*} 
	\begin{center}
		\includegraphics[trim = 0.5cm 0.6cm 0cm 0.4cm, scale = 0.49]{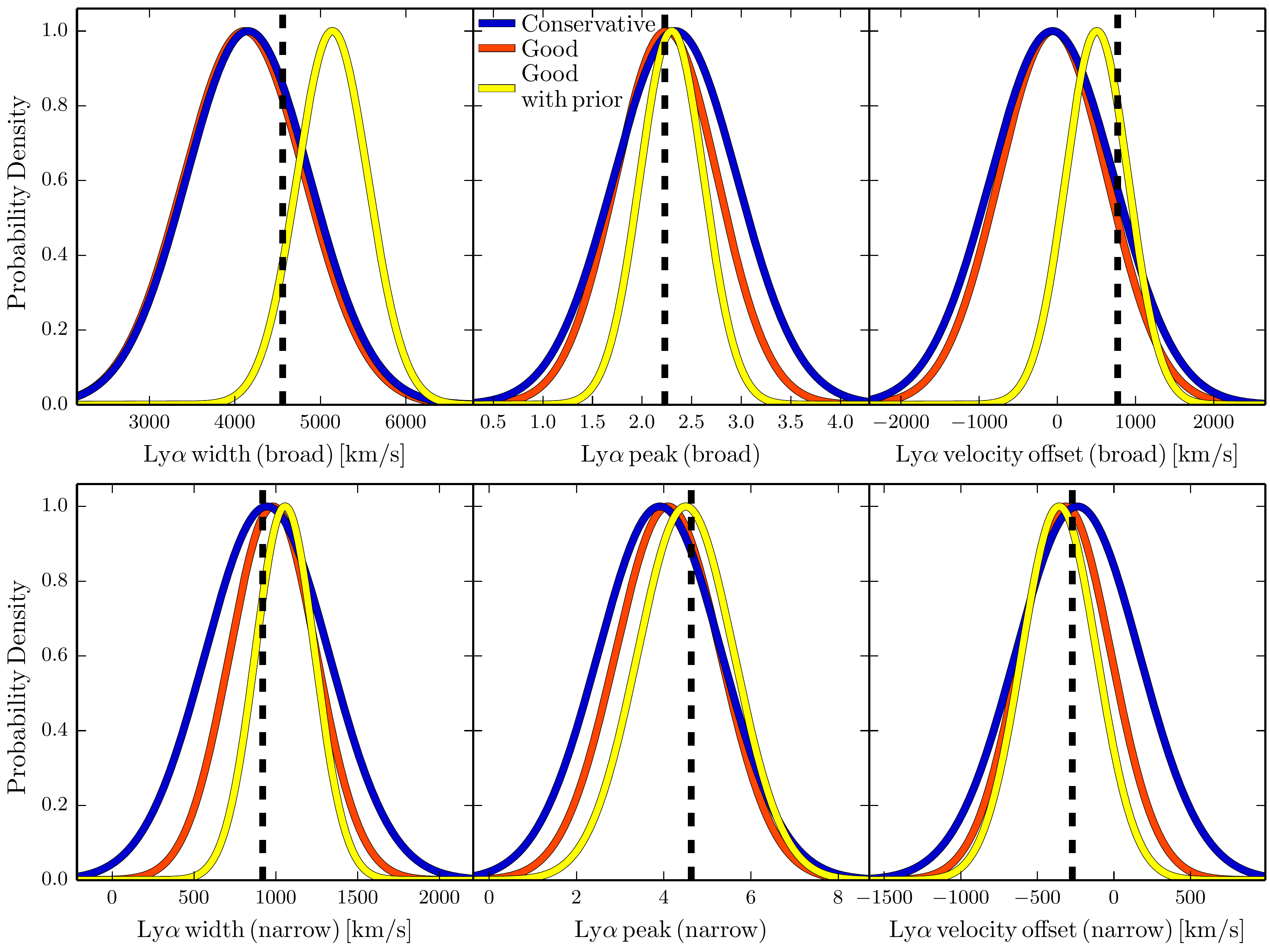}
	\end{center}
\caption[]{The recovered 1D marginalised PDFs for each of the six reconstructed \lya{} emission line parameters obtained after applying the reconstruction method outlined in Section~\ref{sec:Reconstruction} to the example QSO in Figure~\ref{fig:QSOexample}. The vertical dashed lines correspond to the MCMC maximum likelihood fit to the full QSO spectrum, whereas the blue and red curves correspond to the recovered 1D PDFs obtained from using the covariance matrix constructed from the `conservative' and `good' QSO samples respectively. The yellow curve corresponds to reconstructing the \lya{} profile parameters using the `good' sample while in addition applying a prior on the QSO flux within the range $1230 < \lambda < 1275$\AA~(see Section~\ref{sec:fluxprior} for further details). That is, we enforce our reconstructed \lya{} emission line profile to fit the observed spectrum within this region. Note that the peak amplitude is normalised by $f_{1450}$ and is therefore dimensionless. Importantly, we are interested in the joint probability, i.e.\ the full \lya{} line profile. As shown below, the reconstruction performs considerably better than can be inferred from these marginalised PDFs.}
\label{fig:PDFs_Recon}
\end{figure*}

We now present an example to highlight the performance of our approach. In order to do this, we choose the same QSO we showed in Figure~\ref{fig:QSOexample}, fitting for $\lambda > 1275$~\AA\,and recovering the six dimensional estimate for the \lya{} line profile. In restricting our fitting algorithm to $\lambda > 1275$~\AA\, we are not accessing all the information that was used by the full fit to estimate the QSO continuum. While, this does not affect the recovery of the \lya{} peak profile itself, the \lya{} profile plus continuum could be affected. In Appendix~\ref{sec:continuum_comparison} we test this assumption, finding that the QSO continuum parameters can be recovered equivalently from these two approaches, with a small amount of scatter in the QSO spectral index. 

\begin{figure*} 
	\begin{center}
		\includegraphics[trim = 0.4cm 1cm 0cm 0.5cm, scale = 0.492]{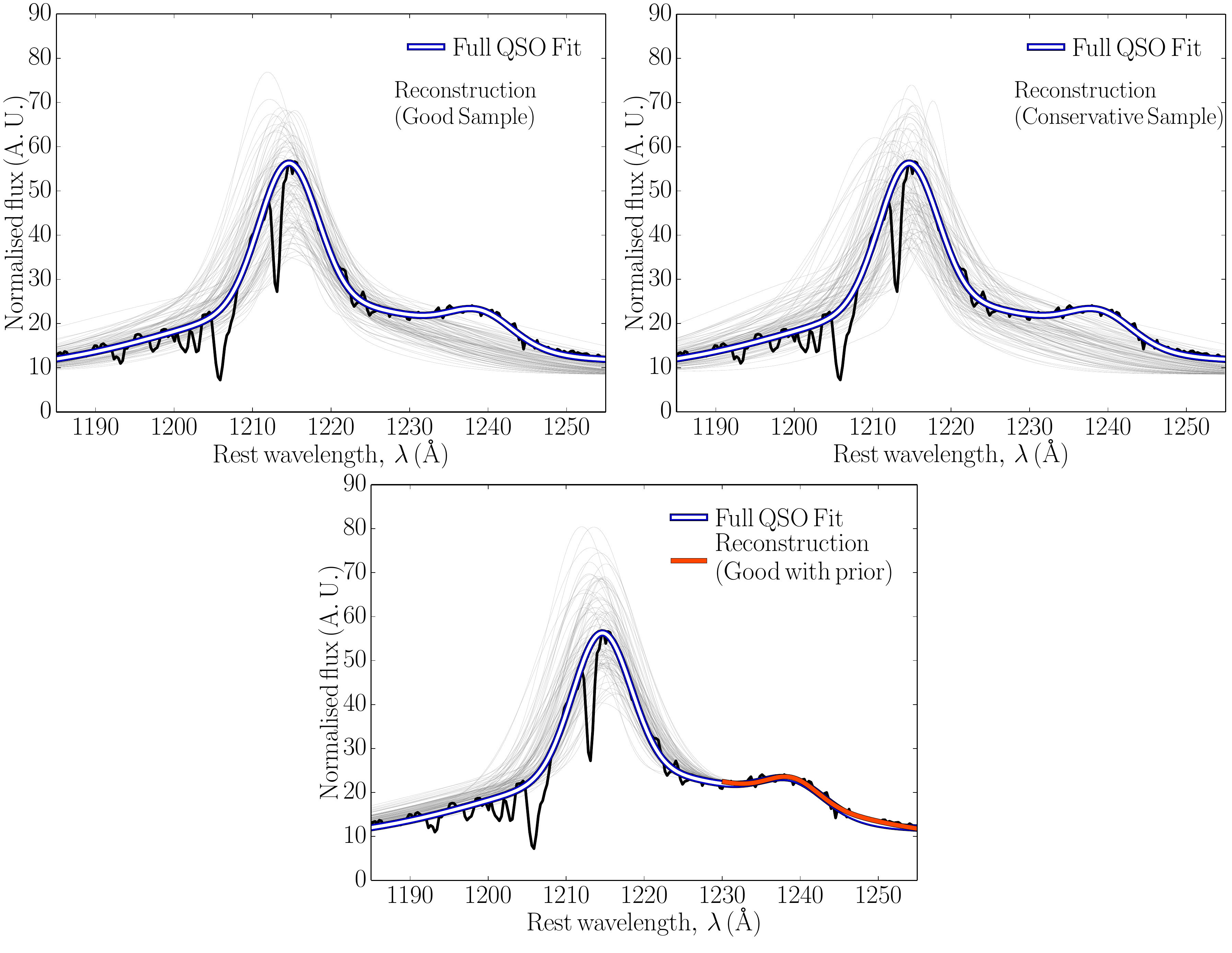}
	\end{center}
\caption[]{A zoom in of the recovered \lya{} emission line profile from our reconstruction procedure. The thin grey curves denote 100 \lya{} line profiles extracted from the reconstructed six-dimensional \lya{} likelihood function. These curves are randomly selected to represent the full posterior distribution for the reconstructed \lya{} profiles, highlighting the relative scale of the errors. The white curves represent the original full MCMC fit of the same QSO (see Figure~\ref{fig:QSOexample}). Black curves are the observed flux of the original QSO spectrum. Note, in these figures, we provide only the emission line component of the full fit (i.e. absorption features are identified and fit, but not shown in the figure). \textit{Top left:} The reconstructed profile from the `good' sample. \textit{Top right:} The reconstructed profile from the `conservative' sample. Note that in both the \textit{top left} and \textit{top right} panels the \nv{} emission feature is not fit in the reconstruction procedure. \textit{Bottom:} The reconstructed profile from the `good' sample utilising the flux prior applied within the range $1230 < \lambda < 1275$\AA~(see Section~\ref{sec:fluxprior} for further details).}
\label{fig:Profile_Recon}
\end{figure*}

Before providing the full reconstructed profile, we first recover the individual marginalised 1D PDFs for each of the six \lya{} emission line parameters to better visualise the relative size of the errors.  In order to obtain the recovered 1D PDFs for the \lya{} profile parameters, we marginalise the six dimensional likelihood function over the remaining five \lya{} parameters. In Figure~\ref{fig:PDFs_Recon} we present these recovered 1D marginalised PDFs, showing in the top row the \lya{} broad line component and in the bottom row, the narrow line component. The vertical black dashed lines represent the recovered values from fitting the full QSO in Figure~\ref{fig:QSOexample}, and the coloured curves represent the recovered 1D marginalised PDFs for each \lya{} parameter for the `good' sample (red) and `conservative' sample (blue). It is clear that both QSO samples recover almost identical best-fit values for each \lya{} parameter, highlighting that both samples are characterised by the same correlations within the covariance matrix. Furthermore, the choice in constructing a `good' sample is more evident here as consistently the `good' sample provides marginally narrower constraints. For the most part, the six reconstructed \lya{} parameters are recovered within the 68 percentile limits of the original fit to the QSO, with the exception being the velocity offset of the broad \lya{} component which is only slightly beyond this limit. Otherwise, the five remaining \lya{} parameters are effectively centred around the expected value from the full fit to the same QSO (Figure~\ref{fig:QSOexample}) which should enable a relatively robust recovery of the full \lya{} line profile.

In Figure~\ref{fig:Profile_Recon} we provide the full reconstructed \lya{} line profile. In the top left panel we show the reconstructed \lya{} line profile obtained from the `good' QSO sample, whereas in the top right panel is the reconstructed profile from the `conservative QSO sample. In all figures, we present 100 reconstructed \lya{} profiles denoted by the thin grey curves which are randomly drawn from the full posterior distribution. This small subset of reconstructed profiles highlight the relative scale of the variations in the total \lya{} line profile peak height, width and position.
The black curve is the raw data from the observed QSO whereas the white curve is the original fit to the full QSO spectrum as shown in Figure~\ref{fig:QSOexample}.

In both, we find the total shape of the reconstructed \lya{} line profile to match extremely well with the original full fit to the same QSO, highlighting the strength and utility of this covariance matrix reconstruction method. Both the `good' and `conservative' samples recover almost the identical reconstructed profile, though the `conservative' QSO sample provides slightly broader errors. Note that for both these QSO samples, the reconstructed \lya{} line profile is systematically below the original fit to the same QSO, indicated by the original QSO fit (white curve) being above the highest density of line profiles\footnote{This is more prevalent when averaging over the full distribution of reconstructed \lya{} profiles.}. However, this systematic offset is only minor (less than 10 per cent in the normalised flux), well within the errors of the reconstruction. Referring back to Figure~\ref{fig:PDFs_Recon}, we can see that this underestimation appears due to the narrow component of the \lya{} peak amplitude (bottom, central panel). 

\subsection{Improving the reconstruction with priors} \label{sec:fluxprior}

We presented in Figure~\ref{fig:Profile_Recon} our best-fit reconstructed \lya{} profiles to a representative QSO drawn from our full sample. However, note that in this reconstruction method we have only used information from the QSO spectrum above $\lambda > 1275$\AA. In doing this, for the case of our example QSO we found our maximum likelihood estimates to slightly underestimate (by less than 10 per cent) the original \lya{} line profile. At the same time, the total 68 per cent marginalised likelihoods of the reconstructed profiles are relatively broad. Motivated by this, we investigate whether we can provide an additional prior on the \lya{} reconstruction profile to further improve the robustness of the recovered profile and to reduce the relative scatter.

In the top panels of Figure~\ref{fig:Profile_Recon}, redward of \lya{} we see that the reconstructed \lya{} line profile drops well below the observed flux of the original QSO. Given we are only reconstructing the \lya{} line, this is to be expected as we are not recovering or fitting the \nv{} emission line. However, in both \lya{} obscured and high-$z$ QSOs, the \nv{} emission line should be relatively unobscured, therefore, the observed flux within this region could be used as a relative prior on the overall \lya{} flux amplitude. 

We therefore include a flux prior\footnote{Note that this choice in phrasing does not imply a `prior' in the Bayesian statistical sense, rather it is used to highlight that we are including additional information into our reconstruction procedure compared to the reconstruction method discussed in Section~\ref{sec:Reconstruction}.} into our \lya{} reconstruction by performing the following steps:
\begin{itemize}
\item As before, we fit the QSO at $\lambda > 1275$\AA, recovering the QSO continuum and all emission line profiles necessary for our covariance matrix approach.
\item Using these estimates, we collapse the 18-dimensional covariance matrix into a six dimensional estimate of the intrinsic \lya{} emission line profile.
\item We then jointly MCMC sample the observed QSO spectrum within the range $1230 < \lambda < 1275$\AA. We fit the \nv{} and \siii{} lines at the same time sampling from our six dimensional reconstructed \lya{} likelihood function obtained from the $\lambda > 1275$\AA\,fit. Fitting to the observed QSO flux, and using the observed noise in the spectrum, we obtain a maximum likelihood for the reconstructed profile. In other words, we require the reconstructed \lya{} line profiles to fit the observed spectrum over the range $1230 < \lambda < 1275$\AA.
\end{itemize}

Implementing this prior on the observed flux closer to $\lambda = 1230$\AA\footnote{This choice of 1230\AA\ is purely arbitrary, and is chosen based on the assumption that a damped absorption signal would not extend this far redward of \lya{}. However, this choice is flexible, and can be adjusted on a case-by-case basis if evidence for stronger attenuation beyond 1230\AA\ is present.}~accesses additional information on the \lya{} profile that was not available through our original $\lambda > 1275$\AA~reconstruction method. Near $\lambda = 1230$\AA, there should be a contribution from the \lya{} broad line component, which is somewhat degenerate with the \nv{} line (as can be seen in Figure~\ref{fig:QSOexample}). By simultaneously fitting the \nv{} line and the \lya{} likelihood function, we use this additional information to place a prior on the \lya{} broad line component, which should then reduce the overall scatter in the six dimensional \lya{} likelihood function.

Referring back to Figure~\ref{fig:PDFs_Recon} we provide an example of this prior applied to the same QSO fit and reconstructed previously. In Figure~\ref{fig:PDFs_Recon}, the yellow curves represent the recovered 1D marginalised PDFs for each of the six \lya{} emission line parameters. Immediately, it is clear that the application of this prior further reduces the relative error in the recovered PDFs. Furthermore, these PDFs remain centred on the originally recovered values, highlighting we have not biased our reconstruction method.

\begin{figure*} 
	\begin{center}
		\includegraphics[trim = 0.3cm 0.7cm 0cm 0.5cm, scale = 0.495]{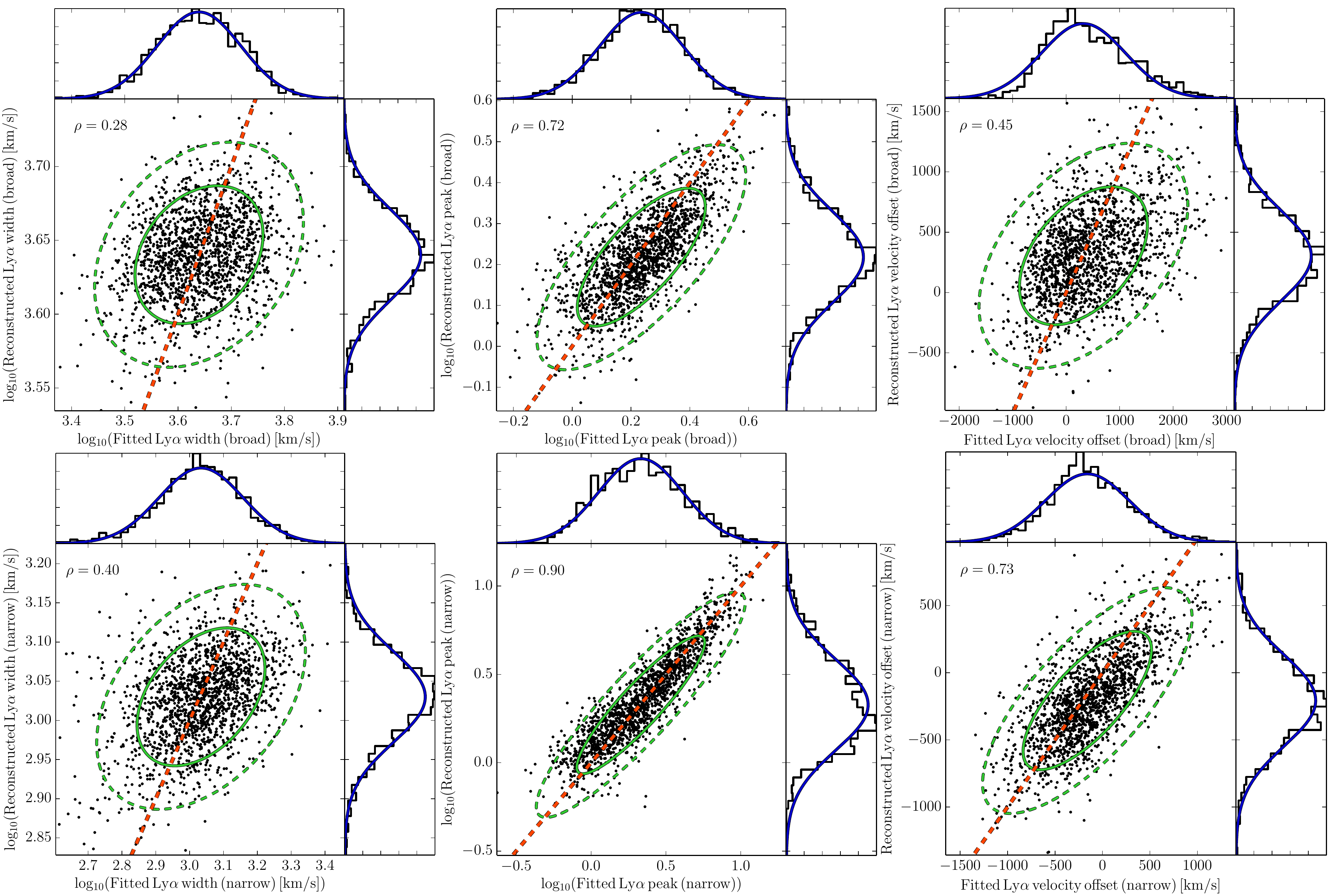}
	\end{center}
\caption[]{A visual characterisation of the performance of the \lya{} emission line reconstruction pipeline at recovering each of the six \lya{} line parameters. We provide the 2D scatter between the full MCMC fits to the QSOs, compared to the reconstructed values from a fit to the same QSO masking the \lya{} region (flux masked at $\lambda < 1275$\AA) and applying our additional prior on the QSO flux near $\lambda = 1230$\AA. Green solid and dashed contours enclose the 68 and 95 per cent scatter of the reconstructed \lya{} parameters relative to their expected (true) value, while the red dashed curves correspond to the one-to-one line on which all points would lie if the reconstruction procedure worked perfectly. Histograms (black curves) correspond to the 1D PDFs of the \lya{} parameters from the full MCMC fitting (top) and the recovered estimate from the reconstruction pipeline fit to the \lya{} masked QSO (right). Blue curves represent the associated Gaussian distribution with equivalent scatter. Note that the peak amplitude is normalised by $f_{1450}$ and is therefore dimensionless.}
\label{fig:Recon_All}
\end{figure*}

In the bottom panel of Figure~\ref{fig:Profile_Recon} we present the full reconstructed \lya{} profile after the addition of this prior. The red curve indicates the fit to the QSO within $1230 < \lambda < 1275$\AA, which we have used as our prior to improve the reconstruction of the \lya{} line profile. At $\lambda < 1230$\AA, we then have the same 100 thin grey curves representing the full posterior distribution of the reconstructed \lya{} likelihood profiles. By applying this additional prior on the total observed flux, we reduce the overall scatter in the reconstructed profile. The maximum likelihood profiles (thin grey curves) now provide a more robust match to the observed QSO.

\subsection{Statistical performance of the reconstruction pipeline}

Thus far we have only applied our reconstruction pipeline to recover the \lya{} line profile from a single, test example selectively drawn from our `good' sample of 1673 QSOs. In order to statistically characterise the performance of the \lya{} reconstruction pipeline across the QSO sample, in Figure~\ref{fig:Recon_All} we present 2D scatter plots of the maximum a posteriori (MAP) estimates\footnote{Note that throughout this work, the recovered MAP estimates from the full posterior distribution do not differ significantly from the peak of the associated marginalised PDFs.} of the reconstructed \lya{} emission line parameters compared to the originally obtained values from the full MCMC fit to the QSO (without masking out the \lya{} line). For each of the six \lya{} emission line parameters, we highlight the 68 and 95 percentiles of the joint marginalised likelihoods for the distributions by the green solid and dashed contours, respectively. Additionally, the red dashed curve demarcates the one-to-one line, along which all QSOs would sit if the reconstruction profile worked idealistically. 

Across the six \lya{} panels, we find strong agreement ($\rho > 0.7$) amongst half of our \lya{} line parameters, those being the peak amplitudes of both the \lya{} broad and narrow components and the velocity offset for the \lya{} narrow line component. For the remainder of the parameters we find moderate to weaker recovery of the original line parameters. However, note that in order to keep this figure as clear as possible we are only providing the MAP estimates. Within Figures~\ref{fig:PDFs_Recon} and~\ref{fig:Profile_Recon} we found that the relative scatter on the reconstructed \lya{} line profile parameters was notable, and therefore the majority of the reconstructed parameters are within the 68 per cent marginalised errors. 

\begin{figure*} 
	\begin{center}
		\includegraphics[trim = 0.3cm 0.7cm 0cm 0.5cm, scale = 0.58]{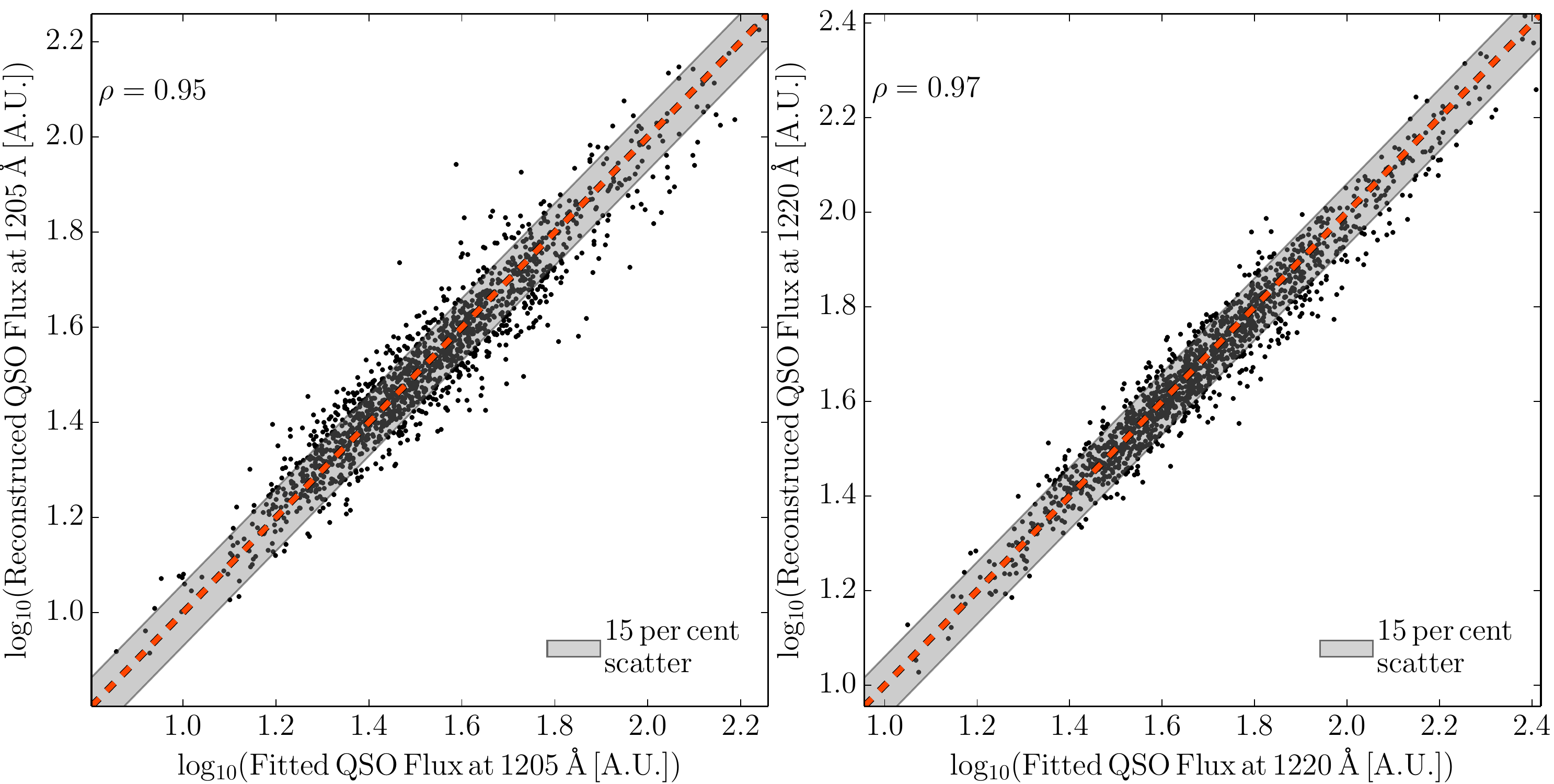}
	\end{center}
\caption[]{A comparison of the maximum likelihood reconstructed \lya{} line flux (including the flux prior; Section~\ref{sec:fluxprior}) to the actual \lya{} line flux obtained from the full fit at a specific wavelength for each QSO within our `good' sample (1673 QSOs; we define 1 A. U. = $10^{-17}\,{\rm erg\,cm^{-2}\,s^{-1}\,\AA^{-1}}$). \textit{Left panel:} blueward of \lya{} ($\lambda=1205$\AA). \textit{Right panel:} redward ($\lambda=1220$\AA). The red dashed curve corresponds to the one-to-one relation, and the grey shaded region encompasses 15 per cent scatter in the reconstructed \lya{} flux relative to the actual measured \lya{} line flux. At $\lambda=1220$\AA~we find the reconstructed \lya{} line flux to be within 15 per cent of the actual \lya{} line flux in $\sim90$~per cent of all QSOs, decreasing to $\sim85$~per cent at $\lambda=1205$\AA.}
\label{fig:FluxDistribution}
\end{figure*}

The reconstructed \lya{} profile parameters highlighted here reflect the correlations recovered from the covariance matrix in Figure~\ref{fig:CovarianceMatrix}. We found strong correlations in the peak amplitudes of the \lya{} profiles, and the narrow component velocity offset. The lack of a strong correlation for the width of the \lya{} line in our covariance matrix, translates to weaker recovery of these parameters. In principle, these weaker correlations could be further strengthened by adding an appropriate prior on the line widths motivated by the statistical distributions recovered from the full sample, or other line properties such as correlations between the equivalent widths.

In Figure~\ref{fig:Recon_All}, there is also slight evidence for a bias in the reconstructed parameters, as highlighted by the orientation of the green contours (68 and 95 percentiles of the reconstructed parameter distributions) relative to the reference one-to-one line. 
However, this could artificially arise as the increase/decrease in any one of these \lya{} line parameters could be compensated for by respective changes in others (i.e.\ model degeneracies), whereas the full six dimensional \lya{} likelihood function takes these model degeneracies into account when estimating the full reconstructed \lya{} profile.

In order to better illustrate the full reconstruction of the joint \lya{} parameter likelihoods, in Figure~\ref{fig:FluxDistribution} we show the information from the six individual \lya{} line parameters as a single, total measured \lya{} line flux at two arbitrarily defined locations blue (1205\AA) and redward (1220\AA) of the \lya{} line centre. We compare the total reconstructed \lya{} line flux against the measured \lya{} line flux from our full fit to the same QSO, providing the reference one-to-one relation as the red dashed curve, and the grey shaded region encompasses the region in which the reconstructed \lya{} line flux is within 15 per cent of the measured flux. Immediately obvious from this figure is that there are no apparent biases in the reconstruction process, i.e.\ we neither systematically over nor underestimate the reconstructed \lya{} line profile. This figure highlights the strength of the \lya{} reconstruction process. Redward of \lya{} (at 1220\AA), closer to our flux prior at 1230\AA, we find that $\sim90$ per cent of all reconstructed \lya{} line profiles have a recovered flux within 15 per cent. As one would expect, the scatter increases blueward of \lya{} (at 1205\AA), however we still find the reconstructed \lya{} line flux to be within 15 per cent for $\sim85$ per cent of our sampled QSOs. This highlights statistically, that the reconstruction process performs an excellent job of recovering the full \lya{} line profile. 

\section{Potential Applications} \label{sec:Discussions}

In this work, we have developed an MCMC fitting algorithm for the sole purpose of characterising the QSO continuum and the emission line profiles within the range $1180{\rm \AA} < \lambda < 2300{\rm \AA}$. Our goal was the construction of a covariance matrix to reconstruct the \lya{} line profile. However, due to the flexibility of the MCMC approach, many other applications could benefit from such a pipeline.

Firstly, we were only interested in correlations amongst the strongest, high ionisation emission line parameters for our covariance matrix. However, various properties of QSOs can be extracted from accurate recovery of the line widths and ratios. For example, the QSO metallicity has been estimated from measuring the \nv{}/\civ{}, \nv{}/He\,{\scriptsize II} and the \sioiv{}/\civ{} line ratios in samples of QSOs from \citet{Nagao:2006p4776} ($2 < z < 4.5$) and \citet{Juarez:2009p4775} ($4 < z < 6.4$). Using the same line ratios, we could recover estimates for the metallicities for all QSOs within our measured sample. 

Several other emission line ratios (e.g.\ the R23 parameter, [O\,{\scriptsize III}]/[O\,{\scriptsize II}], \civ{}/He\,{\scriptsize II}, N-based ratios) can additionally be used as proxies for QSO metallicity \citep[e.g][]{Nagao:2006,Matsuoka:2009,Batra:2014}. By extending our QSO MCMC framework, many other emission lines can be obtained and characterised to improve the QSO metallicity estimates. Crucially, this MCMC approach, would enable a large dataset of QSOs to be rapidly explored. The extension to measuring the metallicities of the QSOs would enable quantities such as the SMBH mass to be recovered through existing correlations between the emission line FWHMs.

In Section~\ref{sec:absorb} we have outlined our method to identify and fit absorption features (see e.g.~\citealt{Zhu:2013} for a more robust approach). While in this work these have been considered contaminants, the prevalence and measurement of these lines could be used to infer properties of the metallicities within intervening absorbers \citep[e.g][]{RyanWeber:2006,Becker:2009,DOdorico:2013} and to reveal the presence of massive outflows of ionised gas from their nuclei \citep[e.g.][and reference therein]{Crenshaw:2003}. For example, we have thrown out any QSOs with strong intervening absorption from systems such as DLAs, however, analysing these sources with our MCMC fitting algorithm could yield measurements on the internal properties of these absorption systems.

Assuming the input quasars to our MCMC are representative of the quasar population as a whole, and further that the UV spectral properties of quasars do not evolve with redshift, our model can be used to predict the intrinsic distribution of quasar spectra at redshifts beyond those from which the model was calibrated ($z\sim2.5$). A similar procedure has been often employed to characterise the colour selection efficiency of quasar surveys \citep[e.g][]{Fan:1999}, although the correlations obtained through our MCMC approach provide more detailed reconstruction of quasar emission features and hence more reliable colour models. Particularly since the \lya{} line plays such a key role in the selection of high-$z$ quasars, our model could be used to identify quasars missing from current surveys due to selection effects and provide more robust statistics for high-$z$ quasar luminosity functions.

In addition, by recovering an estimate for the intrinsic \lya{} emission line profile one can investigate the QSO proximity effect. This approach requires an estimate of the intrinsic QSO luminosity, coupled with the modelling of the \lya{} forest. Within the sphere of influence of the QSO, the photoionisation background is higher than the mean background permeating the IGM. Modelling the transition between the mean IGM background and the drop-off in QSO luminosity has been used by several authors to recover estimates of the photoionisation background in the IGM \citep[e.g][]{Bolton:2005p6088,Bolton:2007p3273,Calverley:2011}. At the redshifts where these studies have been performed (e.g. $2 < z < 6$), the \lya{} line profile is not obscured or attenuated by a neutral IGM. Therefore, one could in principle push the flux prior we used in this work much closer to the \lya{} line centre, to substantially reduce the errors on the intrinsic \lya{} emission line profile blueward of \lya{}.

At $z>6$, for an increasingly neutral IGM, the intrinsic \lya{} emission line can become increasingly attenuated by the Gunn-Peterson IGM damping wing. While existing methods have been developed to access information on the red-side of \lya{} \citep[e.g.][]{Kramer:2009p920}, for the $z=7.1$ QSO ULASJ1120+0641 \citep{Mortlock:2011p1049}, evidence suggests that the IGM damping wing imprint extends further redward \citep[e.g][]{Mortlock:2011p1049,Bolton:2011p1063}, limiting the effectiveness of these approaches. The approach developed in this work should be unaffected by this as here we do not fit the \lya{} line, our reconstruction method therefore is perfectly suited for exploring the potential imprint of the IGM damping wing on the $z=7.1$ QSO \citep[e.g.][]{Greig:2016p1}, along with other future $z>6$ QSOs.

\section{Conclusion} \label{sec:Conclusions}

Characterising the continuum and emission lines properties of QSOs provides a wealth of information on the internal properties of the AGN, such as the mass of the SMBH, the QSO metallicity, star formation rates of the host galaxy, nuclear outflows and winds etc. Furthermore, the intrinsic \lya{} line shape can be used to probe properties of the IGM, such as the mean photoionisation background and the abundance of neutral hydrogen in the IGM at $z>6$.

Motivated by correlations amongst QSO emission lines \citep[e.g.][]{Boroson:1992p4641,Sulentic:2000,Shen:2011p4583}, in this work, we developed a new reconstruction method to recover the intrinsic \lya{} profile. This method is based on the construction of a covariance matrix built from a large sample of moderate-$z$ ($2.0 < z < 2.5$), high S/N (S/N $> 15$) QSOs from the BOSS observational programme. We use this moderate-$z$ sample to characterise the intrinsic \lya{} line profile, where it should be relatively unaffected by intervening neutral hydrogen in the IGM. In order to characterise each QSO within our sample we developed an MCMC fitting algorithm to jointly fit the QSO continuum, the emission lines and any absorption features that could contaminate or bias the fitting of the QSO. We modelled the QSO continuum as a single two parameter power-law ($\propto \lambda^{\alpha_{\lambda}}$), and each emission line is modelled as a Gaussian defined by three parameters, its width, peak amplitude and velocity offset from systemic.

We constructed our covariance matrix from a refined sample of 1673 QSOs, using the high ionisation emission lines \lya{}, \sioiv{}, \civ{} and \ciii{}. Owing to the flexibility of the MCMC framework, we explored various combinations of single and double component Gaussians to characterise these lines. For \lya{} and \civ{} we settled on a double component Gaussian, to model the presence of a broad and narrow component. For the remaining two lines, we considered a single component. This resulted in an $18\times18$ covariance matrix, which we used to investigate new and existing correlations amongst the line profiles. We identified several strong trends from our covariance matrix, most notably the strong positive correlation between the peak amplitudes of the \lya{} and \civ{} narrow components ($\rho=0.8$) and the strong anti-correlation between the \lya{} narrow component peak amplitude and the width of the \ciii{} line ($\rho=-0.74$). These two were the strongest examples of a consistent trend of a positive correlation in the peak amplitudes across all the emission line species and the anti-correlation between the peak amplitudes and line widths.

Using this covariance matrix, we constructed an $N$-dimensional Gaussian likelihood function from which we are able to recover our reconstructed \lya{} line profile. The reconstruction method works as follows:
\begin{itemize}
\item Fit a QSO with our MCMC pipeline within the range $1275{\rm \AA} < \lambda < 2300$\AA, recovering the parameters defining the QSO continuum, and the \sioiv{}, \civ{} and \ciii{} lines.
\item Obtain a six dimensional estimate of the reconstructed \lya{} line profile (modelled as a two component Gaussian) which provides the best-fit profile and correlated uncertainties, by evaluating the $N$-dimensional likelihood function describing our full covariance matrix including a prior on the reconstructed \lya{} line using the observed QSO flux within the range $1230 < \lambda < 1275$\AA.
\end{itemize}
To visually demonstrate the performance of this reconstruction method, we applied it to a randomly selected QSO from our full data set. Finally, we quantitatively assessed its performance by applying it to the full QSO sample, and compared the reconstructed \lya{} profile parameters to those recovered from the original full MCMC fit of the same QSO. We found that estimates for both the \lya{} peak amplitudes are recovered strongly, as is the velocity offset of the narrow line component. For both of the line widths and the broad component velocity offset we find moderate agreement. We additionally explored the total reconstructed \lya{} flux (rather than individual parameters) relative to the original full MCMC fit at two distinct wavelengths blueward (1205\AA) and redward (1220\AA) of \lya{}. Our reconstruction method recovered the \lya{} line flux to within 15 per cent of the measured flux at 1205\AA~(1220\AA)~$\sim85$ ($\sim90$) per cent of the time.

There are several potential applications for both the MCMC fitting method and the \lya{} reconstruction pipeline. The MCMC fitting could be easily modified to measure any emission or absorption feature within a QSO spectrum. With this, many properties of the source QSO could be extracted, for example QSO metallicities. The ability to reconstruct the intrinsic \lya{} line profile could have important cosmological consequences such as improving estimates of the IGM photoionisation background or recovering estimates of the IGM neutral fraction.

\section*{Acknowledgments}

We thank the anonymous referee for their helpful suggestions. AM and BG acknowledge funding support from the European Research Council (ERC) under the European Union's Horizon 2020 research and innovation programme (grant agreement No 638809 -- AIDA -- PI: AM). ZH is supported by NASA grant NNX15AB19G.

Funding for SDSS-III has been provided by the Alfred P. Sloan Foundation, the Participating Institutions, the National Science Foundation, and the U.S. Department of Energy Office of Science. The SDSS-III web site is http://www.sdss3.org/.

SDSS-III is managed by the Astrophysical Research Consortium for the Participating Institutions of the SDSS-III Collaboration including the University of Arizona, the Brazilian Participation Group, Brookhaven National Laboratory, Carnegie Mellon University, University of Florida, the French Participation Group, the German Participation Group, Harvard University, the Instituto de Astrofisica de Canarias, the Michigan State/Notre Dame/JINA Participation Group, Johns Hopkins University, Lawrence Berkeley National Laboratory, Max Planck Institute for Astrophysics, Max Planck Institute for Extraterrestrial Physics, New Mexico State University, New York University, Ohio State University, Pennsylvania State University, University of Portsmouth, Princeton University, the Spanish Participation Group, University of Tokyo, University of Utah, Vanderbilt University, University of Virginia, University of Washington, and Yale University.

\bibliography{Papers}

\appendix

\section[]{Selecting the systemic redshift for our BOSS sample} \label{sec:systemic_redshift}

The reconstruction method discussed within this work is strongly dependant on having an accurate systemic redshift for the QSO sample. In the absence of knowing the true source redshift, systematic errors in its estimation will filter into the recovered line profile parameters, most notably the velocity offset of the line centre. As a result, this will affect any reported line profile correlations which are necessary for the \lya{} line profile reconstruction. Within this appendix, we discuss our attempts to determine the most robust and accurate redshift to be adopted within this work.

The BOSS DR12Q database \citep{Paris:2016p1} provides six available redshift estimates; a visually inspected redshift, $z_{\rm VI}$, the BOSS pipeline redshift based on a decomposed eigenvalue training set, $z_{\rm pipeline}$ (see e.g.\ \citealt{Bolton:2012SDSS}), an automated PCA based approach, $z_{\rm PCA}$, and three other estimates determined from the location of the peak emission for the individual line profiles, $z_{\civ{}}$, $z_{\ciii{}}$ and $z_{\mgii{}}$. Using the earlier DR9Q database, \citet{Font-Ribera:2013} performed a detailed analysis on each of these redshift estimates. These authors found $z_{\rm PCA}$ to be the optimal redshift estimator, returning both the smallest dispersion and bias amongst these six available measurements.

\begin{figure}
	\begin{center}
		\includegraphics[trim = 0.4cm 0.7cm 0cm 0.5cm, scale = 0.57]{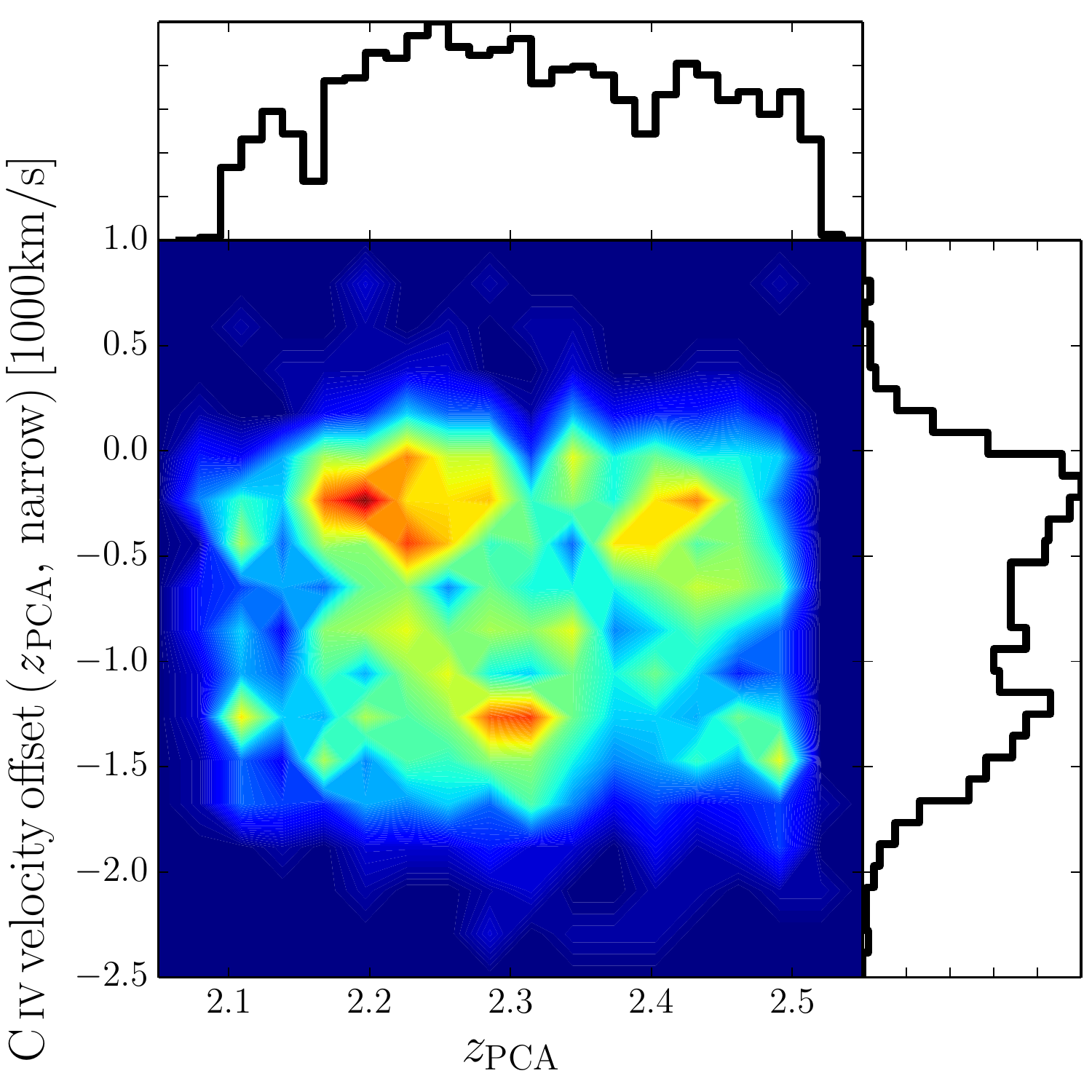}
	\end{center}
\caption[]{Distribution of the velocity offset from the \civ{} narrow component against the estimated PCA redshift, $z_{\rm PCA}$ for our `good' sample of 1673 QSOs. In the central panel is a two-dimensional histogram normalised to unity, with red denoting the highest density of QSOs. In the top (right) panels are the one-dimensional histograms for the distribution of $z_{\rm PCA}$ (\civ{} narrow line component velocity offset).}
\label{fig:z_PCA}
\end{figure}

\begin{figure}
	\begin{center}
		\includegraphics[trim = 0.4cm 0.7cm 0cm 0.5cm, scale = 0.57]{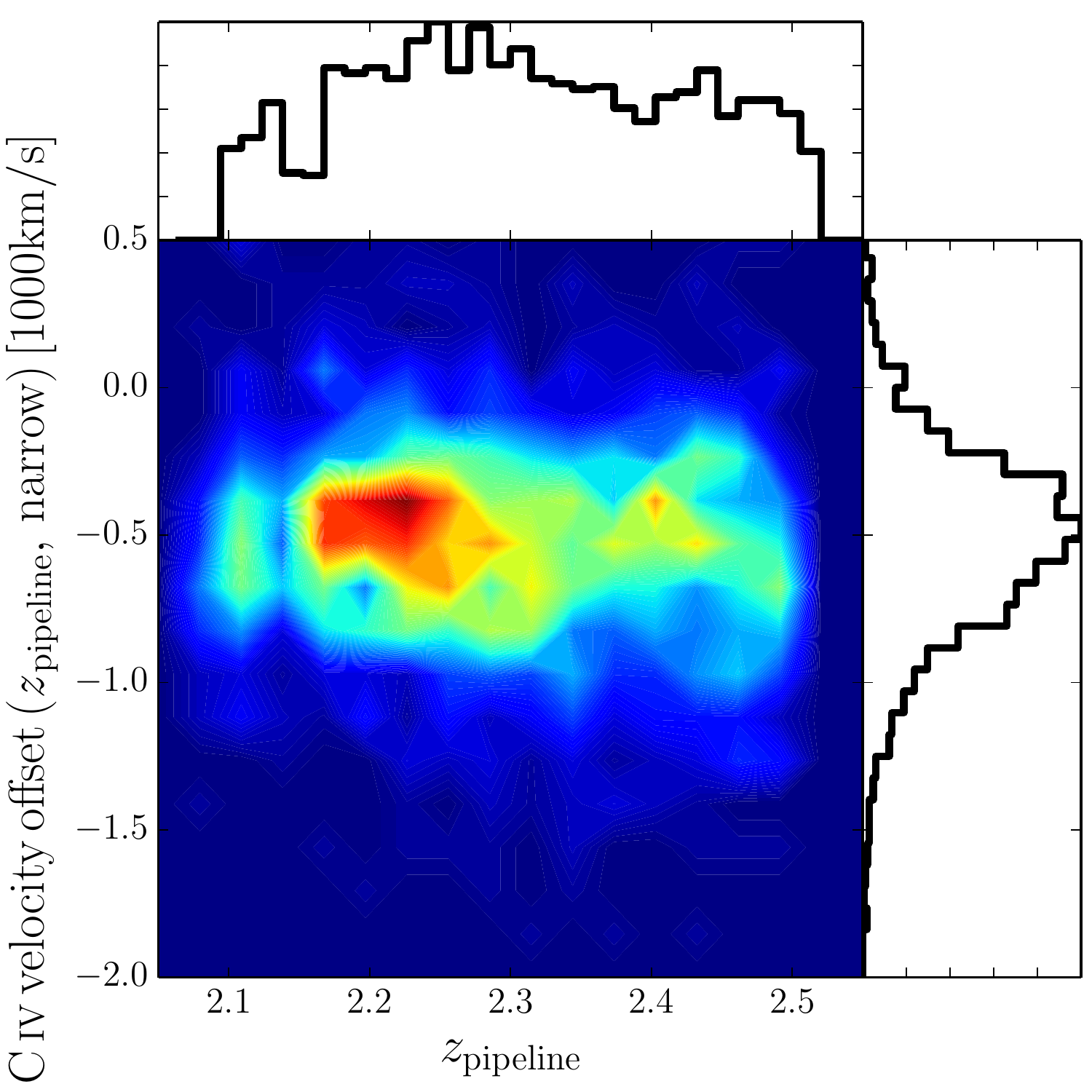}
	\end{center}
\caption[]{Same as Figure~\ref{fig:z_PCA}, except showing the \civ{} narrow component measured from the $z_{\rm pipeline}$ redshift estimator.}
\label{fig:z_pipe}
\end{figure}

However, in adopting $z_{\rm PCA}$ as the optimal redshift estimator we find a bias within our selected sample of QSOs. In Figure~\ref{fig:z_PCA}, we show the velocity offset of the narrow component of the \civ{} line (measured with $z_{\rm PCA}$) as a function of the estimated redshift from the PCA approach using our `good' sample of 1673 QSOs. In the right most panel, we show the one-dimensional histogram for the narrow component of \civ{}. Immediately obvious from this figure, is the non-Gaussian (bi-modal) nature of the distribution. Contrast this with the equivalent figure of the narrow component of \civ{} for the $z_{\rm pipeline}$ redshift estimator shown in Figure~\ref{fig:z_pipe}, which shows no obvious departure from Gaussianity. We have additionally tested this for the other redshift estimators (e.g.\ $z_{\rm VI}$ and $z_{\mgii{}}$) and found no evidence for the same bi-modal distribution as shown by $z_{\rm PCA}$. 

In the central panel of Figure~\ref{fig:z_PCA}, we show the two-dimensional histogram, with red signifying the highest density of QSOs. Presenting the data in this manner immediately highlights the source of error. The two peaks in the \civ{} velocity offset distribution are driven by two separate, discrete regions within the \civ{}--$z_{\rm PCA}$ parameter space: the primary peak in the \civ{} velocity offset at $\sim-250$~km/s (for all redshifts except the `hole' at $z\sim2.3$) and the secondary peak at $\sim-1250$~km/s (at $z\sim2.3$). This discrete `overdensity' of QSOs at $z\sim2.3$ is most likely an unphysical artefact. The likely cause of this discrete transition near $z\sim2.3$ is the redshifting of the \mgii{} line between strong OH sky line features. Typically, the \mgii{} emission line is one of the strongest anchors in which the source redshift can be determined. At $2.25 \lesssim z \lesssim 2.33$, \mgii{} is observed within a region that has a relatively low sky background with few sky lines, resulting in a higher \mgii{} S/N. However, at $z\lesssim2.25$ and $z\gtrsim2.33$, the \mgii{} feature appears amongst OH forest regions with very strong sky lines, resulting both in a noisier signal and it being strongly affected by systematics owing to the sky line residuals. Therefore, the weakening of the \mgii{} information in determining the QSO redshift within the PCA approach at $z\gtrsim2.33$ is the likely cause of this transition. 
Due to this bi-modal nature, which appears to be intrinsic only to the $z_{\rm PCA}$ redshift estimator, and since throughout this work we have assumed the distributions of line profiles can all be represented by a Gaussian distribution, we refrain from adopting the PCA redshift within this analysis.

The next three best candidates after $z_{\rm PCA}$ from the \citet{Font-Ribera:2013} analysis are $z_{\rm VI}$, $z_{\rm pipeline}$ and $z_{\mgii{}}$\footnote{We do not consider either of $z_{\civ{}}$ or $z_{\ciii{}}$ as these have the largest dispersion and bias.}. Of these three, $z_{\mgii{}}$ has the smallest bias, followed by $z_{\rm pipeline}$ and $z_{\rm VI}$. In contrast, $z_{\rm VI}$ has the smallest dispersion, followed by $z_{\rm pipeline}$ and $z_{\mgii{}}$. In this work, we do not consider using the visually inspected redshift as: (i) it has the strongest systematic bias of these choices (ii) it is the least objective measure (not entirely automated), requiring intervention to improve the redshift.

The QSOs within our `good' sample are selected within $2.08 \leq z \leq 2.5$. By adopting, $z_{\mgii{}}$ instead of $z_{\rm pipeline}$, we lose all QSOs above $z>2.4$, where no $z_{\mgii{}}$ redshift is reported for any of the sources in DR12Q. Ultimately, the loss of sources reduces the total number of QSOs within our sample by $\sim25$ per cent. Despite the reduction in sample size, the $z_{\mgii{}}$ and $z_{\rm pipeline}$ samples return almost identical covariance matrices, signifying consistency in both redshift estimates. The major difference between the two samples is the notably stronger correlation in the velocity offsets for the emission line parameters within the $z_{\mgii{}}$. However, this is to be expected owing to the differences in how the redshifts are estimated. For $z_{\mgii{}}$, a single line profile is used to estimate the redshift, whereas in $z_{\rm pipeline}$ several line profiles are compared to a training set to estimate the best candidate redshift. By using several line profiles to determine the source redshift, intrinsic scatter within the source spectrum can cause the redshift estimates to deviate compared to those from a single line measurement such as $z_{\mgii{}}$. Despite $z_{\mgii{}}$ providing the stronger line profile components for the velocity offsets than the $z_{\rm pipeline}$, we refrain from using the \mgii{} redshifts as these correlations may be artificially high owing to their estimation from only a single line profile. We deem the correlations from $z_{\rm pipeline}$ sample to be more conservative, while also benefitting from the addition of maintaining a larger statistical sample.

\begin{figure}
	\begin{center}
		\includegraphics[trim = 0.4cm 0.7cm 0cm 0.5cm, scale = 0.57]{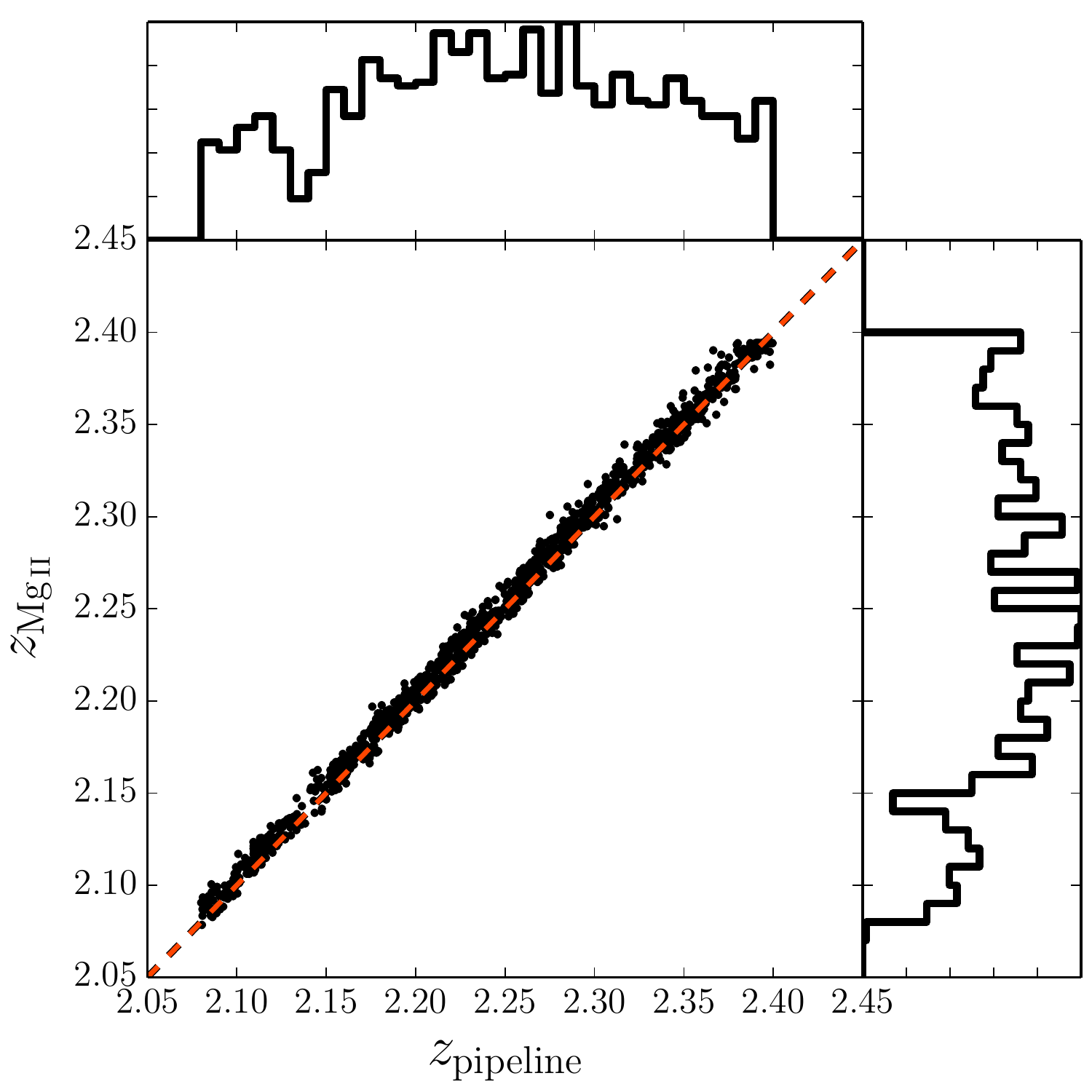}
	\end{center}
\caption[]{The distribution of 1274 QSOs drawn from our `good' sample which contain both a \mgii{} and BOSS pipeline redshift. Top and left panels are histograms of the BOSS pipeline redshift, $z_{\rm pipeline}$ and \mgii{} derived redshift, $z_{\mgii{}}$. The red dashed curve denotes the one-to-one relation, highlighting a slight offset between the two redshift estimates.}
\label{fig:z_offset}
\end{figure}

Finally, $z_{\mgii{}}$ and $z_{\rm pipeline}$ return relatively similar dispersions, with $z_{\mgii{}}$ being less biased. With the vast majority of the QSOs in our sample having both redshifts, in principle we can attempt to calibrate our $z_{\rm pipeline}$ sample by the $z_{\mgii{}}$ sample. In Figure~\ref{fig:z_offset} we provide the $z_{\mgii{}}$--$z_{\rm pipeline}$ parameter space, with the red dashed curve denoting the one-to-one relation, from which one can see a small offset. We recalibrate our sample for this offset by creating a pipeline corrected redshift, 
\begin{eqnarray}
z_{\rm pipeline\_corr} = z_{\rm pipeline} + \langle z_{\mgii{}} - z_{\rm pipeline} \rangle,
\end{eqnarray}
for which we recover a small offset of $\sim2.77\times10^{-3}$~($\sim361$~km/s at $z\sim2.3$).

Employing this BOSS pipeline corrected redshift, we perform the full analysis as performed within Sections~\ref{sec:Covariance} and~\ref{sec:Recon}. With respect to the covariance matrices, we recover effectively identical correlations amongst our line profile parameters. In order to better gauge the performance of these two possible samples, we recover the fraction of QSOs with a reconstructed flux recovered to within 15 per cent at both 1205\AA\ and 1220\AA\ (see e.g.\ Figure~\ref{fig:FluxDistribution}). Though the differences are minor, we find that the original BOSS pipeline redshift ($z_{\rm pipeline}$) performs marginally better than $z_{\rm pipeline\_corr}$. At 1205\AA, we recover the QSO flux to within 85 (82) per cent for $z_{\rm pipeline}$ ($z_{\rm pipeline\_corr}$) and 90 (88) per cent at 1220\AA\footnote{For completeness, for $z_{\mgii{}}$ we recover the QSO flux to within 15 per cent for 80 per cent of the QSOs at 1205\AA\ and 85 per cent at 1220\AA.}. Given that the primary aim of this work is to reconstruct \lya{} emission profiles, then the most natural choice for a redshift estimate is the one which performs best at this reconstruction; the line parameter correlations are a secondary product. Therefore, we conclude that owing to the marginally improved statistics on the recovery of the QSO flux from the $z_{\rm pipeline}$ sample relative to $z_{\rm pipeline\_corr}$, and the reasons mentioned previously for not considering other redshift estimates, the best redshift to adopt within this work is $z_{\rm pipeline}$.

\section[]{Preference for the number of Gaussian components for a line profile} \label{sec:line_component}

In the construction of our covariance matrix in Section~\ref{sec:covariance}, we model the \lya{} and \civ{} emission lines as a double component Gaussian whereas for the \sioiv{} and \ciii{} lines we consider only a single component Gaussian. In this appendix we detail quantitatively the reasoning for our choices. To objectively differentiate between the various combinations we use the Bayes Information Criterion (BIC; \citealt{Schwarz:1978p1,Liddle:2004p5730}). The BIC can be computed using,
\begin{eqnarray} \label{eq:BIC}
{\rm BIC} = -2\,{\rm ln}(\mathcal{L}) + k\,{\rm ln}(N),
\end{eqnarray}
where $\mathcal{L}$ is recovered from the $\chi^{2}$ fit to the individual emission line profile, $k$ is the number of free parameters and $N$ is the number of available data points. A lower recovered BIC for an individual fit to the emission line profile is considered to be a better characterisation of the line.

The comparison between two possible sets of parameters to describe the emission line profile can be quantitatively described by the difference in the two BICs, $\Delta$BIC. A recovered $\Delta$BIC $> 10$, provides very strong evidence for the model with the lower BIC, whereas for $6 < \Delta$BIC $< 10$ strong evidence exists. Throughout this appendix we discuss each emission line profile individually, and recover the $\Delta$BIC for each model to distinguish between our preferred choices.

\begin{table*}
\caption{We use the Bayes Information Criterion (BIC) to distinguish whether the emission lines used within our covariance matrix should be fit with a single or a double component Gaussian. For each potential combination we present the $\Delta$BIC to distinguish between very strong evidence ($\Delta$BIC $>10$) and strong evidence ($6 < \Delta$BIC $< 10$) for a certain combination of lines (first column) relative to an alternative combination (second) column. See text for further details regarding the various combinations of line profiles considered.}
\begin{tabular}{@{}lccccccc}
\hline
Line fitting type & Line fitting type & QSOs with $\Delta$BIC $ > 10$ & QSOs with $6 < \Delta$BIC $< 10$\\
(preferred option) & (compared against) & (per cent) & (per cent) \\
\hline
\lya{} double component, \nv{} and \siii{} & \lya{} double component and \nv{} & 60.9 & 1.1 \\
\lya{} double component, \nv{} and \siii{} & \lya{} single component, \nv{} and \siii{} & 79.4 & 0.7 \\
\lya{} double component, \nv{} and \siii{} & \lya{} single component and \nv{} & 93.0 & 0.4 \\
\hline
\sioiv{} single component & \sioiv{} double component & 64.7 & 6.9 \\
\hline
\civ{} double component & \civ{} single component & 94.0 & 0.3 \\
\hline
\ciii{} single component and \aliii{} & Only \ciii{} double component & 53.8 & 2.9 \\
\ciii{} single component and \aliii{} & \ciii{} double component and \aliii{} & 45.7 & 6.3 \\
\ciii{} single component and \aliii{} & Only \ciii{} single component & 92.2 & 0.9 \\
\hline
\end{tabular}
\label{tab:BIC}
\end{table*}

\subsection[]{The \lya{} profile}

For the \lya{} emission line profile, we consider the full emission line complex within the rest-frame wavelength range $\lambda = [1170,1280]$~\AA. This provides three different potential line species to be fit, \lya{}, \nv{} and the \siii{} line. Given the importance of the \nv{} line for the full line complex, we always fit the \nv{} line in addition to the \lya{} line. Therefore, we consider four different combinations to potentially characterise the full \lya{} line complex. Firstly, we consider the \lya{} line as a double component Gaussian, with a single component Gaussian each for the \nv{} and \siii{} emission lines. Next, the same, but only a single component \lya{} component. Finally, we consider these same two combinations excluding the \siii{} line.

In Table~\ref{tab:BIC}, we report the $\Delta$BIC for these four combinations of emission lines. We compute the BIC as outlined in Equation~\ref{eq:BIC}, recovering the $\chi^{2}$ within the wavelength range $\lambda = [1170,1280]$~\AA. In the construction of Table~\ref{tab:BIC}, we use the `conservative' sample consisting of 2653 QSOs. We then recover the $\Delta$BIC from the difference between a model with a double component Gaussian for \lya{} and the single components for \nv{} and \siii{}. It is clear from this table that this is the preferred choice for modelling the \lya{} line complex. Importantly, we find strong evidence for a double component \lya{} emission line in over 80 per cent of the QSOs we consider. In principle, we could have explored whether further components for the \lya{} line are preferred, however, for the sake of simplicity we have refrained from doing so. Furthermore, while the \siii{} emission line is typically only a weak ionisation line, and in some cases not resolved within individual QSOs, we still find strong evidence for its inclusion in the QSO fitting (preferred in $\sim62$ per cent of our sample).

\subsection[]{The \sioiv{} profile}

Unlike the \lya{} line considered above, the blended \sioiv{} doublet is a strong, isolated emission line. We therefore consider the simplified choice of whether a single or double component Gaussian is preferred for this line within the wavelength range $\lambda = [1360,1440]$~\AA. From Table~\ref{tab:BIC}, we find that a single component Gaussian is very strongly preferred at $\sim65$ per cent, which becomes $\sim72$ per cent when the criteria is lowered to only strong evidence. 

\subsection[]{The \civ{} profile}

For the \civ{} doublet, we consider the emission line complex within the wavelength range, $\lambda = [1500,1600]$~\AA. Again, the \civ{} emission line is an isolated, very strong high ionisation line, therefore we only need to consider whether a single or double component Gaussian is preferred. In Table~\ref{tab:BIC}, we recover very strong evidence for the choice of a double component Gaussian, with this being preferred 94 per cent of the time. 

\subsection[]{The \ciii{} profile}

The \ciii{} emission line is not an isolated line, instead it is the strongest line in a complicated emission line profile \citep[e.g.][]{VandenBerk:2001p3887}. Surrounding the \ciii{} line are contributions from the Fe pseudo-continuum, as well as contributions from the \aliii{} and Si\,{\scriptsize III]} line. In this work we consider the contribution of the Fe pseudo-continuum to be negligible to the \ciii{} line, and for all intents and purposes we additionally consider the Si\,{\scriptsize III]} contribution negligible. This latter point arises from the insufficient S/N of these BOSS QSOs to fully resolve and separate out this line contribution. While it may be present in the highest S/N QSOs in our sample, for the vast majority it cannot be easily distinguished. Furthermore, from \citet{VandenBerk:2001p3887}, the relative line fluxes differ by over a factor of a hundred.

We consider four scenarios to characterise the \ciii{} line complex. A double component \ciii{} line in combination with a single component Gaussian for \aliii{}. A single component \ciii{} and \aliii{} line, and then just a single or double component Gaussian for the \ciii{} line, without considering the \aliii{} line. In Table~\ref{tab:BIC}, we find just a single \ciii{} emission line to be very strongly disfavoured with respect to a single component \ciii{} and \aliii{} ($\sim92$ per cent). However, in the case of the remaining scenarios it is not as clear.

Comparing the single \ciii{} and \aliii{} components to a purely double component \ciii{} (no \aliii{}), we find marginally stronger evidence for the single \ciii{} scenario (preferred $\sim54$ per cent of the time). However, note that in allowing for a double component \ciii{}, these two scenarios are effectively the same. For example, the broad component can just fit the \aliii{} line, leaving a single \ciii{} line. This degeneracy therefore tends to draw these two scenarios to be more equally preferred. As a result, the single \ciii{} and \aliii{} scenario is considered to be preferable.

In the case of comparing a single or double component \ciii{} line in combination with \aliii{}, we find comparable results for the $\Delta$BIC. Between these two scenarios we find a single component very strongly preferred in $\sim46$ per cent of the QSOs sampled, which increases to $\sim52$ per cent if we lower the criteria to just strong evidence. It is therefore, not immediately clear from the $\Delta$BIC which is the preferred fitting scenario to consider for the \ciii{} line complex. As a result, we choose to consider a single \ciii{} component with a \aliii{} contribution purely on the grounds of there being less model parameters, removing potential degeneracies between a broad and narrow component \ciii{} from our covariance matrix.

\section[]{QSO quality assessment examples} \label{sec:QSO_QA}

\begin{figure*}
	\begin{center}
		\includegraphics[trim = 0cm 0.7cm 0cm 0.5cm, scale = 0.49]{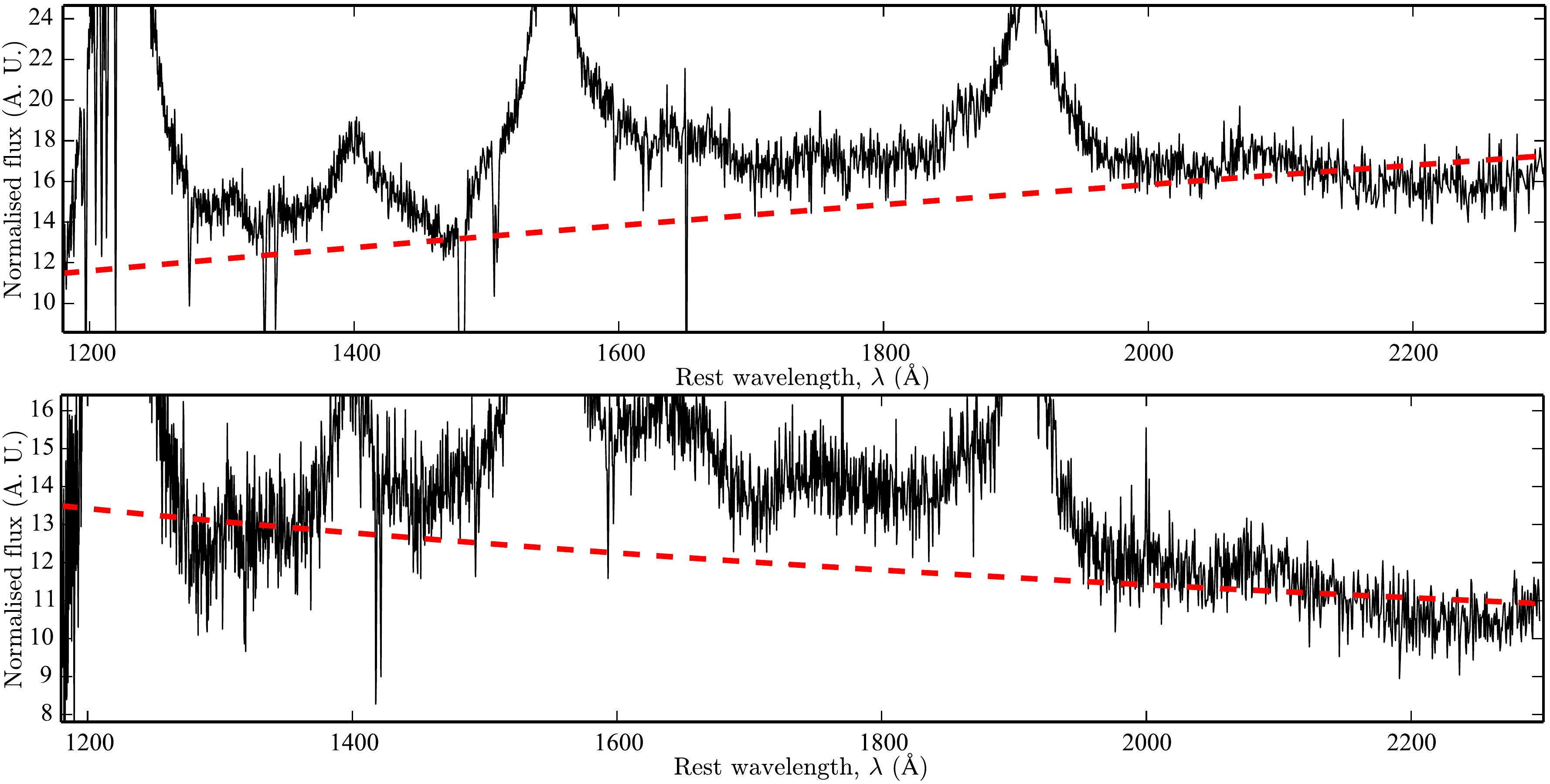}
	\end{center}
\caption[]{Representative examples of QSOs removed from our samples due to continuum estimation issues.  \textit{Top panel:} The recovery of a positive QSO continuum spectral index ($f_{\lambda} \propto \lambda^{\alpha_{\lambda}}$). A small number of QSOs contain a positive spectral index, counter to the vast majority of QSOs which have a negative spectral index. \textit{Bottom panel:} A QSO continuum where a single power-law continuum cannot fit the observed spectrum. In this example, a two component continuum with a break near \civ{} would improve the fit. Only a rare few QSOs show this behaviour, therefore there is no motivation to retain them. Note, 1 A. U. = $10^{-17}\,{\rm erg\,cm^{-2}\,s^{-1}\,\AA^{-1}}$.}
\label{fig:QA_RemovedContinuum}
\end{figure*}

In Section~\ref{sec:QA}, we briefly outlined the visual quality assessment we performed to obtain the two QSO samples for the covariance matrix. Within this appendix we provide a more detailed explanation of the respective criteria we consider, as well as providing a select few examples which best indicate the various decisions which we have made.

\subsection{Removal of QSOs}

The first step of our quality assessment was the removal of any QSOs from which we recovered a poor characterisation of the QSO continuum. In Figure~\ref{fig:QA_RemovedContinuum} we provide two examples of QSOs which were removed failing this criteria. In the top panel, the recovered QSO returns a positive spectral index, $\alpha_{\lambda}$, which is counter to the negative spectral indices recovered by the vast majority of our QSO sample. The recovery of a positive spectral index is unphysical, and likely arises from flux calibration errors. Only a few QSOs within our sample are found to have a positive spectral index. In the bottom panel of Figure~\ref{fig:QA_RemovedContinuum}, we provide an example of a QSO where a single power-law is insufficient to characterise the QSO continuum within our fitting region. To adequately characterise this QSO, a double component continuum pivoting on \civ{} would be required. The rarity in recovering such an object likely points to this being an error in the data reduction pipeline, or the flux calibration.

\begin{figure*} 
	\begin{center}
		\includegraphics[trim = 0.5cm 1.5cm 0cm 0.8cm, scale = 0.285]{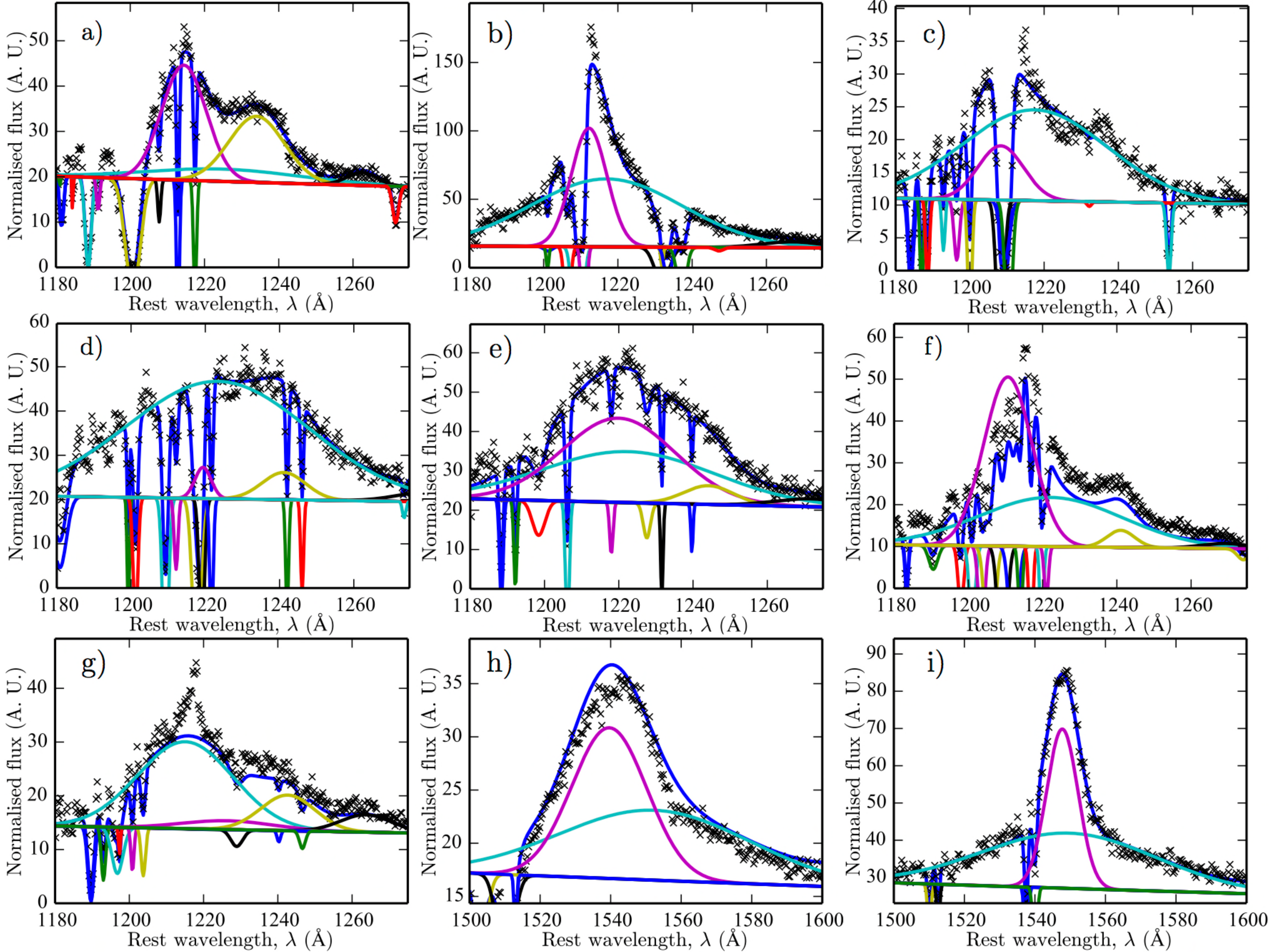}
	\end{center}
\caption[]{Representative examples of QSOs removed from our samples due to a variety of issues. \textit{Top left:} Intervening strong \lya{} absorber. \textit{Top centre:} Strong \lya{} absorption offset from line centre impacting the intrinsic profile recovery. \textit{Top right:} Strong \lya{} absorption at line centre. \textit{Middle left:} A substantial number of \lya{} absorption features contaminating the \lya{} emission line profile. \textit{Middle centre:} No prominent \lya{} emission profile. \textit{Middle right:} Significant \lya{} line absorption which becomes degenerate with the \lya{} emission line profile. \textit{Bottom left:} The MCMC algorithm is unable to fit the intrinsic \lya{} profile owing to absorption impacting the \lya{} emission line profile. \textit{Bottom centre:} \civ{} absorption in the line wing, impacting the broad component determination. \textit{Bottom right:} Small \civ{} absorption feature which artificially prefers a narrower line component than should be recovered.}
\label{fig:QA_RemovedLya}
\end{figure*}

\begin{figure*} 
	\begin{center}
		\includegraphics[trim = 0.5cm 0.8cm 0cm 0.6cm, scale = 0.49]{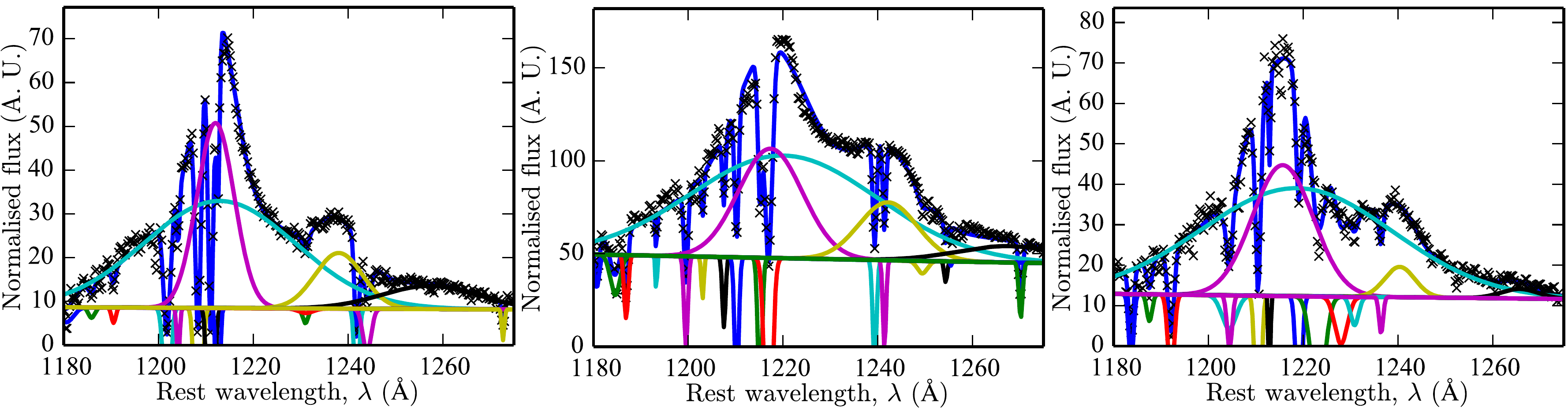}
	\end{center}
\caption[]{Representative examples of \lya{} profiles which are classified in the `conservative' sample (more uncertainty in the recovery of the intrinsic \lya{} profile). \textit{Left panel:} A large number of clustered absorption lines blue-ward of \lya{} which are identified and well fit. The pristine red-side of the \lya{} line should enable the intrinsic \lya{} profile to be well characterised. \textit{Centre panel:} A large number of absorption lines, which are well fit and characterised. The broader \lya{} absorption feature will likely impact the recovery of the intrinsic profile, but there is still sufficient evidence that the intrinsic \lya{} profile might be well estimated. \textit{Right panel:} A large number of absorption features on both the blue and red side of \lya{}. The absence of line centre absorption should enable the intrinsic profile to be recovered with some small uncertainty.}
\label{fig:QA_Conservative}
\end{figure*}

\begin{figure*} 
	\begin{center}
		\includegraphics[trim = 0.5cm 0.8cm 0cm 0.6cm, scale = 0.49]{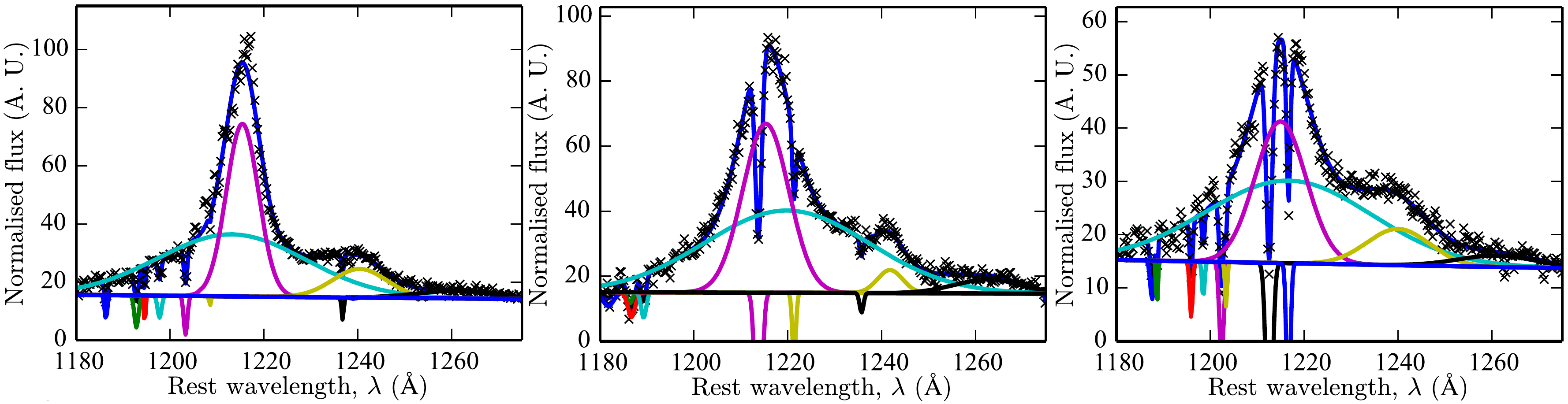}
	\end{center}
\caption[]{Representative examples of \lya{} profiles which are classified in the `good' sample. \textit{Left panel:} The absence of any absorption. \textit{Centre panel:} small absorption, offset from the centre of \lya{}. These features are identified and fit, and therefore do not impact on the ability to fit the intrinsic \lya{} profile. \textit{Right panel:} A larger number of absorption feature, all identified and well fit. The narrow width of the absorption features around \lya{} should not impact the fitting of the \lya{} profile.}
\label{fig:QA_Good}
\end{figure*}

The next step in our visual QSO quality assessment is the removal of any QSO spectra which have either (i) missing sections of the observed QSO flux which overlap with our chosen emission lines (ii) a large number of absorption features (iii) either an intervening dense neutral absorber or sufficiently strong/broad absorption blueward of \civ{} (iv) a \lya{} line region that is not well fit or characterised by our MCMC pipeline. In Figure~\ref{fig:QA_RemovedLya} we present nine examples of QSOs removed from our sample. We discuss each one of these QSOs below\footnote{Note that in Figures~\ref{fig:QA_RemovedLya} -- \ref{fig:QA_Good} we represent the corresponding fitted absorption profiles as being subtracted from the continuum flux level rather than the true flux level. As a result, in some cases it appears that the absorption extends to negative flux. However, this choice is purely to aid the visual representation of the full emission profile fits.}:
\begin{itemize}
\item[\textbf{a)}] A clear, dense neutral absorber in the direct vicinity of the QSO. Presence of the neutral absorber completely removes any ability to recover a \lya{} broad component.
\item[\textbf{b)}] Two strong absorption features near \lya{}. The first, blue ward of \lya{} removes almost half of the \lya{} line. Even in the presence of this absorber, the intrinsic \lya{} line profile recovery is still reasonably good. However, it still under predicts the true profile. Furthermore, absorption on the \nv{} line centre removes any ability to recover the \nv{} line, resulting in a much broader \lya{} component than should be recovered.
\item[\textbf{c)}] Strong \lya{} absorption at the line centre. Without a sufficient fraction of the \lya{} line centre, the intrinsic \lya{} line cannot be recovered. While a broad \lya{} component can still be estimated, no narrow component can be reliably obtained.
\item[\textbf{d)}] Another, even more extreme example of strong absorption on \lya{} line centre. Presence of several strong \lya{} absorption features completely removes any information to obtain the \lya{} narrow line component.
\item[\textbf{e)}] A \lya{} line profile with no prominent narrow line component. As a result, the two Gaussian components both prefer to be broad, and are therefore degenerate. The absorption on line centre likely contaminated the \lya{} peak determination leading to the two preferred degenerate broad components.
\item[\textbf{f)}] A large, connected series of narrow absorption features centred on \lya{}. Since the entire \lya{} profile is contaminated by absorption, measuring the \lya{} narrow component peak amplitude becomes degenerate with the depth of the absorption features.
\item[\textbf{g)}] The MCMC pipeline struggles to recover the extremely narrow, sharp peak around \lya{} line centre. Evidence of two small absorption features (which are not identified by our algorithm) on either side of line centre appear to cause this peak to become too narrow to fit.
\item[\textbf{h)}] Significant blue ward absorption features near the \civ{} line contaminates the recovery of the broad component. Not a BAL QSO, however the absorption features are still sufficient to contaminate the recovery of the full line profile.
\item[\textbf{i)}] Deep, narrow absorption features on the \civ{} line profile. The unfortunate location of these features at the wing of the \civ{} narrow component causes a narrower \civ{} line than should be expected. The depth of these features causes this behaviour, weaker absorption would not impact the narrow component recovery.
\end{itemize}

\subsection{Examples of QSOs in the `conservative' sample}

In Figure~\ref{fig:QA_Conservative} we provide three examples of QSOs which are classified as `conservative' only (not considered for the `good' sample). In the first example (left panel), several strong narrow absorption features are detected and fit. These are offset from the line centre of \lya{}, allowing the peak height of \lya{} to still be reliably estimated. Though the intrinsic profile appears to be well recovered, the uncertainties in the peak amplitude owing to potential degeneracies with the relative strengths of the absorption features and the number of identified absorption features causes this to be only characterised as `conservative'. 

The second example (middle panel) has a strong absorption feature on \lya{} line centre. However, the relative strength of this feature is not sufficient to completely contaminate the peak of the \lya{} line (unlike similar examples in Figure~\ref{fig:QA_RemovedLya}). It appears the full \lya{} line profile is recovered well, despite numerous narrow absorption features and the strong central absorption. The uncertainties surrounding the narrow \lya{} component determination from the strong absorption results in its assessment as `conservative' only. Finally, in the last example (right panel), the full \lya{} line profile is contaminated by a large number of narrow absorption features. None of these are located at line centre, nor do they appear to impact the recovery of the \lya{} line profile. However, the sheer number of features causes this QSO to be classified as conservative.

\subsection{Examples of QSOs in the `good' sample}

Finally, in Figure~\ref{fig:QA_Good} we present three examples of QSOs explicitly selected for our refined `good' sample (all QSOs in the `good' sample appear in the `conservative' sample). In the first example (left panel), there are only a few very small absorption features. In the absence of anything stronger, these QSOs are classified as `good'. The second example (middle panel), once again has very few absorption features. Despite being strong, narrow absorption lines, these are well characterised and do not impact the recovery of the intrinsic \lya{} line at all. In the final example (right panel), a larger number of absorption lines are present. Importantly though, these are located offset from the \lya{} line centre, and therefore we are confident that the true profile is recovered. Note, this profile is similar to the first example in Figure~\ref{fig:QA_Conservative}. The reason this QSO is classified as `good' and the previous was `conservative' arises from the number of absorption lines. In this example, only two narrow features impact the \lya{} line, whereas in the previous case, several lines have resulted in a smaller fraction of the total \lya{} profile being unaffected by the absorption features. Ultimately, this is only a minor difference that does not change any of the conclusions of this work, since in Section~\ref{sec:Bias} we found similar correlation matrices between the two samples.

\section[]{Recovery of continuum parameters from $\lambda>1275{\rm \AA}$} \label{sec:continuum_comparison}

\begin{figure*} 
	\begin{center}
		\includegraphics[trim = 0.5cm 0.7cm 0cm 0.6cm, scale = 0.585]{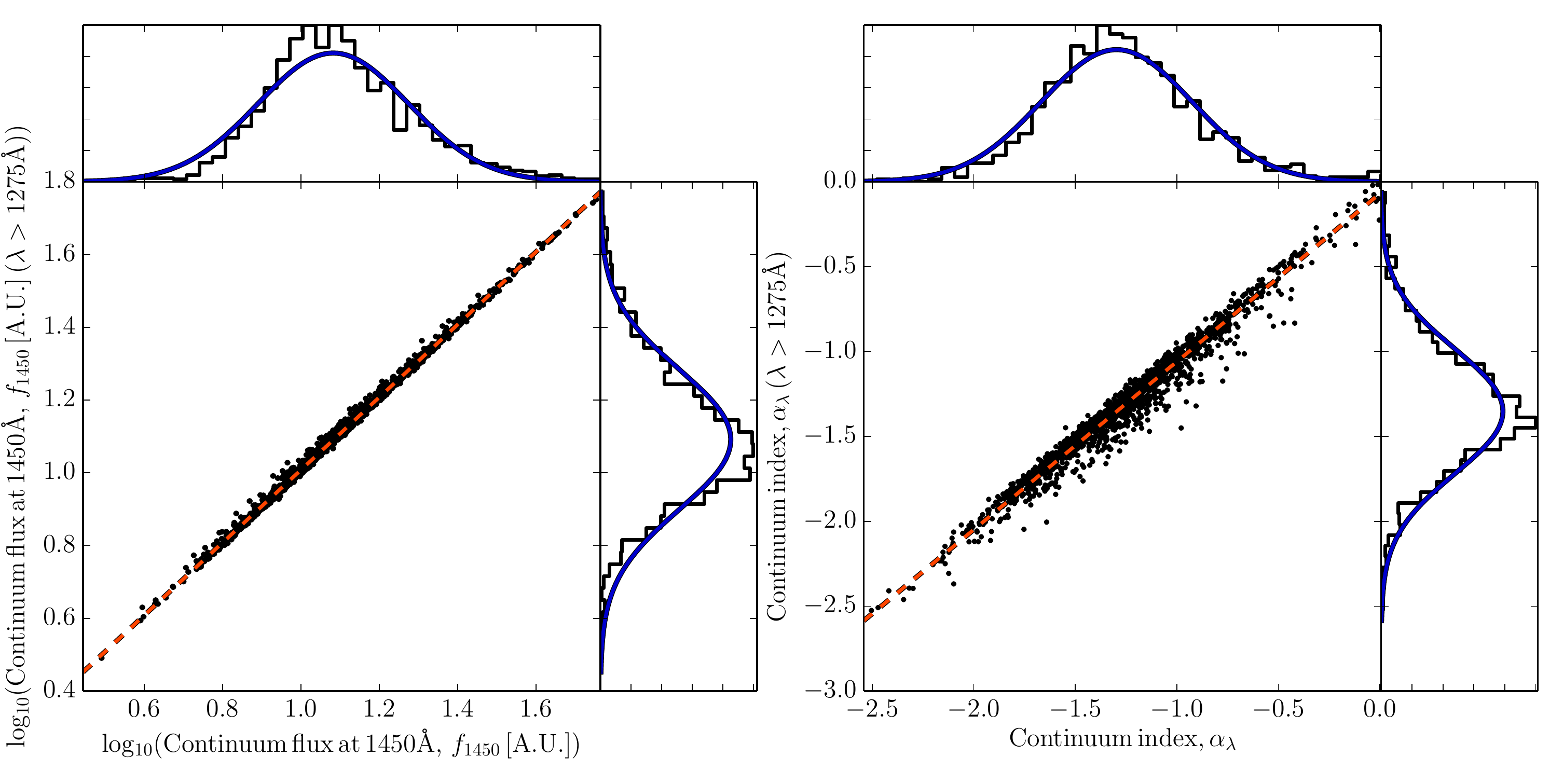}
	\end{center}
\caption[]{2D scatter plots comparing the estimation of the two continuum parameters between the full MCMC fit and the \lya{} masked QSO fit for reconstruction (masked at $\lambda < 1275$\AA). \textit{Left panel:} The continuum flux normalised at 1450\AA, $f_{1450}$. \textit{Right panel:} The continuum spectral index, $\alpha_{\lambda}$. The red dashed curve corresponds to the one-to-one relation. There is very little scatter about the one-to-one relation, indicating there are no biases in the estimation of the continuum parameters when fitting either the full QSO, or fitting at $\lambda > 1275$\AA. Note, 1 A. U. = $10^{-17}\,{\rm erg\,cm^{-2}\,s^{-1}\,\AA^{-1}}$.}
\label{fig:ContinuumComparison}
\end{figure*}

In Section~\ref{sec:ReconExample}, we alluded to an assumption regarding the usage of the QSO continuum between the full MCMC fitting and the reconstruction profile fitting. Specifically, we assume that the continuum measured from a \lya{} obscured or high-$z$ QSO in the region ($\lambda > 1275$\AA) will be equivalent to the continuum recovered from fitting the full QSO within the range $1180{\rm \AA} < \lambda < 2300$\AA. However, by not fitting in the $1180{\rm \AA} < \lambda < 1275$\AA\,region, we could be losing information from the \lya{} line profile (and surrounding region) that could be otherwise used to refine the estimate of the QSO continuum. Of course, this cannot be tested on an actual QSO that requires reconstruction, but instead we can statistically test this assumption across the full QSO sample we have used within this work.

In Figure~\ref{fig:ContinuumComparison}, we present the 2D scatter of the two QSO continuum parameters, the continuum flux amplitude normalised to $1450$~\AA~($f_{1450}$, left panel) and the spectral index, $\alpha_{\lambda}$ (right panel). To facilitate the comparison, the red dashed curve corresponds to the one-to-one relation. Note that while the flux amplitude is normalised at $1450$\AA, which should be unaffected by the different fitting regions, we allow a perturbation on this flux to refine its true normalisation. In the case of this continuum normalisation, we find very little scatter, and all QSOs lie along the one-to-one relation indicative of their being no adverse effects from this assumption.

In the case of the QSO spectral index, we again find all QSOs to essentially lie perfectly along the one-to-one relation. However, in several instances, there are QSOs which deviate away from this line. This small scatter indicates that for a select few QSOs, fitting in the $1180{\rm \AA} < \lambda < 1275$\AA\,region is important for the recovery of the true continuum spectral index. Importantly, around $\lya{}$ even for the extreme QSOs in this 2D plot, the level of discrepancy in the continuum flux will be at most 10 per cent. Furthermore, this is purely for the continuum flux, and not the total line flux. Referring to Figure~\ref{fig:Profile_Recon}, a 10 per cent error on the continuum flux at \lya{} will fall well within the errors of the reconstructed \lya{} emission line profile. Therefore, within the model uncertainties of the \lya{} reconstruction pipeline, this scatter is not significant to drastically affect the reconstructed \lya{} line profile and hence the assumption on using the same QSO continuum is justified.

\end{document}